\documentclass[twocolumn,american,nofootinbib]{revtex4}
\usepackage{ae}
\usepackage{aecompl}
\usepackage[T1]{fontenc}
\usepackage[latin1]{inputenc}
\usepackage{amsmath}
\usepackage{graphicx}
\usepackage{amssymb}

\makeatletter
\usepackage{psfrag}

\usepackage{babel}
\makeatother
\begin{document}

\title{Reheating-volume measure in the landscape }

\author{Sergei Winitzki}

\affiliation{Department of Physics, Ludwig-Maximilians University, Munich, Germany}

\begin{abstract}
I recently proposed the {}``reheating-volume'' (RV) prescription
as a possible solution to the measure problem in {}``multiverse''
cosmology. The goal of this work is to extend the RV measure to scenarios
involving bubble nucleation, such as the string theory landscape.
In the spirit of the RV prescription, I propose to calculate the distribution
of observable quantities in a landscape that is conditioned in probability
to nucleate a finite total number of bubbles to the future of an initial
bubble. A general formula for the relative number of bubbles of different
types can be derived. I show that the RV measure is well-defined and
independent of the choice of the initial bubble type, as long as that
type supports further bubble nucleation. Applying the RV measure to
a generic landscape, I find that the abundance of Boltzmann brains
is always negligibly small compared with the abundance of ordinary
observers in the bubbles of the same type. As an illustration, I present
explicit results for a toy landscape containing four vacuum states
and for landscapes with a single high-energy vacuum and a large number
of low-energy vacua.
\end{abstract}
\maketitle

\section{Introduction and summary\label{sec:Introduction-and-motivation}}

In many cosmological scenarios the fundamental theory does not predict
with certainty the values of observable cosmological parameters, such
as the effective cosmological constant and the masses of elementary
particles. This is the case even for some models of inflation driven
by a scalar field (see e.g.~\cite{Garcia-Bellido:1993wn,Garcia-Bellido:1994vz}
for early work) as well as for the {}``landscape of string theory''~\cite{Bousso:2000xa,Susskind:2003kw,Kachru:2003aw,Douglas:2003um};
see also the {}``recycling universe''~\cite{Garriga:1997ef} models.
In these latter models, the fundamental theory admits a large number
of disjoint vacuum states. Transitions between these states are possible
through bubble nucleation; the interior of a bubble appears as an
infinite homogeneous open universe~\cite{Coleman:1980aw}, if one
disregards the small probability of bubble collisions (see Refs.~\cite{Garriga:2006hw,Bousso:2008as}
for analyses of bubble collisions). The presently observed universe
is situated within a bubble (called a {}``pocket universe'') of
some type. 

A common feature of these cosmological models is the presence of \emph{eternal
inflation}, i.e.~the absence of a global end to inflation in the
entire spacetime (see Refs.~\cite{Linde:1993xx,Guth:2000ka,Winitzki:2006rn}
for reviews). Eternal inflation gives rise to infinitely many causally
disconnected regions of the spacetime where the cosmological observables
may have significantly different values. In the context of the string-theoretic
landscape, eternal inflation entails the nucleation of (potentially)
infinitely many nested bubbles of different vacuum types.

The program outlined in the early works~\cite{Garcia-Bellido:1994ci,Vilenkin:1994ua,Vilenkin:1995yd},
which dealt with eternal inflation of random walk type, was to calculate
the probability distribution of the cosmological parameters as measured
by an observer randomly located in the spacetime. The main diffuculty
in obtaining such probability distributions is due to the infinite
volume of regions where an observer may be located. An eternally inflating
universe contains an infinite, inhomogeneous, and topologically complicated
spacelike hypersurface (the reheating surface) where observers may
be expected to appear with a constant density per unit 3-volume. 

In the landscape scenarios, one encounters a kind of infinity that
is in some sense more ill-behaved than in the random-walk inflationary
scenarios. Not only each pocket universe may contain infinitely many
observers, but also the number of different pocket universes in the
entire spacetime is infinite. Pocket universes of different types
are not statistically equivalent to each other because they have different
rates of nucleation of other pocket universes. There seems to be no
natural ordering on the set of all pocket universes throughout the
spacetime, since most of the pocket universes are spacelike separated.
To emphasize the mutual causal independence of pocket universes, one
calls such a spacetime a {}``multiverse.''

In summary, eternal inflation is a stochastic process that generates
a topologically complicated and noncompact locus of points where observers
may appear. A {}``random location'' of an observer within that locus
is a mathematically undefined concept, similarly to the concept of
an integer number {}``uniformly chosen'' among all the integers,
or a real number {}``uniformly chosen'' among all the reals. This
is the root cause of the technical and conceptual difficulties known
collectively as the {}``measure problem'' in multiverse cosmology
(see Refs.~\cite{Guth:2000ka,Aguirre:2006ak,Winitzki:2006rn,Vilenkin:2006xv,Guth:2007ng,Linde:2007nm}
for reviews). Nevertheless, one may try to formulate a prescription
for calculating probabilities of observer-based events. Such a prescription,
also called a {}``measure,'' should in some sense correspond to
the intuitive notion of probability of observation at a {}``random''
location in the spacetime.

Several measure prescriptions have been proposed in the literature.
The proposals that apply directly to landscape scenarios are the {}``holographic''
measure~\cite{Bousso:2006ev,Bousso:2006ge} (see also the recent
proposal~\cite{Garriga:2008ks}), the {}``comoving horizon cutoff''~\cite{Garriga:2005av,Easther:2005wi,Vanchurin:2006qp},
the {}``stationary measure''~\cite{Linde:2007nm,Clifton:2007bn},
the measure on transitions~\cite{Aguirre:2006na}, and the {}``pseudo-comoving''
measure~\cite{Linde:2006nw,DeSimone:2008bq,Bousso:2008hz,DeSimone:2008if}.
In the absence of a unique definition of the measure, one judges a
cutoff prescription viable if its predictions are not obviously pathological.
Possible pathologies include the dependence on choice of spacetime
coordinates~\cite{Winitzki:1995pg,Linde:1995uf}, the {}``youngness
paradox''~\cite{Linde:1994gy,Vilenkin:1998kr}, and the {}``Boltzmann
brain'' problem~\cite{Dyson:2002pf,Albrecht:2004ke,Page:2006dt,Linde:2006nw,Vilenkin:2006qg,Page:2006ys,Bousso:2007nd,Gott:2008ii}.
Various measures have been used for predicting cosmological parameters,
most notably the cosmological constant, in the landscape scenarios
(see, e.g., Refs.~\cite{Schwartz-Perlov:2006hi,Clifton:2007bn,Bousso:2007er,Bousso:2007kq,SchwartzPerlov:2006hz,Olum:2007yk,SchwartzPerlov:2008he}).

The purpose of this paper is to extend the most recently proposed
{}``reheating-volume'' (RV) measure~\cite{Winitzki:2008yb,Winitzki:2008ph},
originally formulated in the context of random-walk inflation, to
landscape scenarios. The basic idea of the RV proposal is to select
multiverses that are very large but (by rare chance) have a finite
total number of observers. In the context of a string landscape scenario
(or a {}``recycling universe''), this can happen if sufficiently
many anti-de Sitter or Minkowski bubbles nucleate everywhere, collide,
and merge. In such a case, there will be a finite time after which
no de Sitter regions remain and no further nucleations can occur.
Hence, there will be a finite time after which no more observers are
created anywhere. By this construction, one obtains a subensemble
of multiverses having a fixed, finite total number $N_{\text{obs}}$
of observers. These finite multiverses with very large $N_{\text{obs}}$
are regarded as controlled approximations to the actual infinite multiverse.
The observer-weighted statistical distribution of any quantity within
a finite multiverse can be obtained by ordinary counting, since the
total number of observers within any such multiverse is finite. The
limit of that statistical distribution as $N_{\text{obs}}\rightarrow\infty$
is the final result of the RV prescription.

It was shown in Refs.~\cite{Winitzki:2008yb,Winitzki:2008ph} that
the RV measure is gauge-invariant, independent of the initial conditions,
and free of the youngness paradox in the context of random-walk inflation.
Presently I investigate whether the same features persist in an application
of the RV measure to landscape scenarios. In particular, it is important
to obtain RV predictions with respect to the {}``Boltzmann brain''
problem that has been widely discussed.

In principle, the RV prescription can be extended to landscape models
in different ways, depending on the precise choice of the ensemble
of finite multiverses. The ensemble of multiverses with a fixed total
number of observers $N_{\text{obs}}$ (where one counts both the ordinary
observers and the {}``Boltzmann brains'') appears to be the natural
choice. However, it is difficult to compute the number of observers
directly and unambiguously. Instead of the total number of observers,
I propose to fix the total number $n_{\text{tot}}$ of bubbles nucleated
to the future of an initial bubble.

The total number of ordinary observers in bubbles of a given kind
is proportional to the volume of the reheating surface in those bubbles.
It is known that the {}``square bubble'' approximation~\cite{Bousso:2007nd},
which neglects the effects of bubble wall geometry, is adequate for
the purposes of volume counting. Then the evolution of the landscape
is well described by the approximate model called {}``inflation in
a box''~\cite{Aryal:1987vn,Winitzki:2005ya,Winitzki:2008yb}. In
that approximation, one keeps track only of the number of new bubbles
nucleated in previously existing bubbles, and each new bubble is assumed
to be instantaneously nucleated exactly of Hubble size in comoving
coordinates. Motivated by this approximation, in this paper I study
a simplified definition of the RV measure for a landscape scenario
(see Sec.~\ref{sub:Extending-the-RV} for details): One requires
the total number of bubbles of all types, $n_{\text{tot}}$, to be
finite and evaluates the statistical distribution of bubble types
(or other cosmological observables) in the limit $n_{\text{tot}}\rightarrow\infty$.
In principle, this limit can be calculated if the bubble nucleation
rates are known. Since this paper is a first attempt to perform this
techically challenging calculation, I concentrate only on bubble abundances
and on the relative abundance of Boltzmann brains. I neglect the increased
number of observers due to additional slow-roll inflation within bubbles;
this effect was considered in Ref.~\cite{Winitzki:2008yb} and requires
additional complications in the formalism.

A landscape scenario may be specified by enumerating the available
$N$ types of vacua by a label $j$ ($j=1$, ..., $N$) and by giving
the Hubble rates $H_{j}$ within bubbles of type $j$. One can, in
principle, compute the nucleation rate $\Gamma_{j\rightarrow k}$
describing the probability (per unit four-volume of spacetime) of
creating a bubble of type $k$ within bubbles of type $j$.%
\footnote{For some recent work concerning the determination of the bubble nucleation
rates, see Refs.~\cite{Aguirre:2006ap,Podolsky:2008du,Johnson:2008kc,Freivogel:2008wm,Johnson:2008vn}. %
} It is convenient to work with the dimensionless rates,\begin{equation}
\kappa_{j\rightarrow k}\equiv\frac{4\pi}{3}\Gamma_{j\rightarrow k}H_{j}^{-4}.\end{equation}
The rate $\kappa_{j\rightarrow k}$ equals the probability of having
a bubble of type $k$ within a 3-volume of one horizon in a bubble
of type $j$, during a single Hubble time.%
\footnote{The notation $\kappa_{j\rightarrow k}$, chosen here for its visual
clarity, corresponds to $\kappa_{kj}$ of Ref.~\cite{Garriga:2005av}
and to $\Gamma_{jk}$ of Ref.~\cite{Winitzki:2008yb}.%
} Explicit expressions for $\kappa_{j\rightarrow k}$ are available
in some landscape scenarios. In what follows, I assume that $\kappa_{j\rightarrow k}$
are known.

In Sec.~\ref{sub:A-toy-landscape} I apply the RV measure proposal
to a toy model of the landscape with four vacua (the FABI model of
Ref.~\cite{Garriga:2005av}). In this model, the vacua labeled $F$
and $I$ are de Sitter (dS) and the vacua labeled $A$ and $B$ are
anti-de Sitter (AdS) states. One assumes that only the transitions
$F\rightarrow I$, $I\rightarrow F$, $F\rightarrow A$, and $I\rightarrow B$
are allowed, with known nucleation rates $\kappa_{FI}$, $\kappa_{IF}$,
etc., per unit Hubble 4-volume. I show in Eq.~(\ref{eq:ratios 2})
that the RV-regulated bubble abundances depend on the value of the
dimensionless number\begin{equation}
\eta\equiv\left(\frac{\kappa_{IB}}{\kappa_{FA}}\right)^{\nu+1}\frac{\kappa_{FI}}{\kappa_{IF}},\quad\nu\equiv e^{3}.\end{equation}
Here the constant $\nu$, introduced for convenience, is simply the
number of statistically independent Hubble regions after one $e$-folding.
Barring fine-tuned cases, one expects that the value of $\eta$ is
either much larger than 1 or much smaller than 1, since the nucleation
rates may differ by exponentially many orders of magnitude. By relabeling
the vacua ($F\leftrightarrow I$ and $A\leftrightarrow B$) if necessary,
we may assume that $\eta\ll e^{-6}$. Then the bubble abundances are
approximately described by the ratios\begin{equation}
p(I):p(F):p(A):p(B)\approx\frac{1}{\nu^{2}}:\frac{1}{\nu}:1:\left[\frac{\eta}{\nu^{\nu-1}}\right]^{\frac{1}{\nu+1}}.\end{equation}
This result can be interpreted as follows. Each of the $I$ bubbles
produces $\nu$ bubbles of type $F$, and each of the $F$ bubbles
produces $\nu$ bubbles of type $A$. The abundance of $B$ bubbles
is neligible compared with other bubbles. Heuristically, the chain
of transitions $I\rightarrow F\rightarrow A$ can be interpreted as
the {}``dominant'' chain in the landscape. The fine-tuned case,
$e^{-6}<\eta<e^{6}$, is considered separately, and the result is
given by Eq.~(\ref{eq:regime1 ans fine-tuned}).

I then consider the abundance of Boltzmann brains (BBs) in the FABI
landscape (Sec.~\ref{sub:The-Boltzmann-brain}). The total number
of BBs is proportional to the total number of Hubble regions ($H$-regions,
or 4-volumes of order $H^{-4}$) in de Sitter bubbles. The coefficient
of proportionality is the tiny nucleation rate $\Gamma^{BB}$ of Boltzmann
brains, which is of order $\exp(-10^{50})$ or smaller. In comparison,
ordinary observers occur at a rate of at least 1 per horizon volume.
It turns out that (after applying the RV cutoff) the total number
of $H$-regions of dS types $F$ or $I$ is approximately equal to
the total number of nucleated bubbles of the same type. Hence, the
BBs are extremely rare compared with ordinary observers. 

In Sec.~\ref{sec:A-general-landscape} I extend the same calculations
to a general landscape with an arbitrary number of vacua. The RV prescription
predicts a definite ratio $p(j)/p(k)$ between the number of bubbles
of types $j$ and $k$. I derive a formula for the ratio $p(j)/p(k)$
that involves all the parameters of the landscape. With the help of
mathematical results derived in Sections~\ref{sub:Uniqueness-of-the}
and \ref{sub:Asymptotic-behavior-of-lambda}, it is possible to show
in full generality that the RV measure gives well-defined results
that are independent of the initial conditions. Nevertheless, actually
performing the required calculations for an arbitrary landscape remains
a daunting task. To obtain explicit expressions in a semi-realistic
setup, I calculate the bubble abundances for a landscape that contains
a single high-energy vacuum and a large number of low-energy vacua.
The result is an approximate formula {[}Eq.~(\ref{eq:pjk ans example})]
for the ratio $p(j)/p(k)$ expressed directly through the nucleation
rates of the landscape.

I also demonstrate in Sec.~\ref{sub:Boltzmann-brains-landscape}
that the abundance of Boltzmann brains is negligible compared with
the abundance of ordinary observers in the same bubble type. 

To conclude, the present work demonstrates that the RV measure has
attractive features and may be considered a viable candidate for the
solution of the measure problem in multiverse cosmology. More work
is needed to investigate the dependence of the predictions on the
precise details of the definition of the ensemble $E_{n}$. I have
developed an extensive mathematical framework for the calculations
in the RV prescription and obtained first results for specific landscapes.
However, a more powerful approximation scheme is desirable so that
the predictions of the RV measure can be more easily obtained for
landscapes of general type. Ultimately, the viability of the RV measure
is to be judged by its predictions for cosmological observables in
realistic landscapes. These issues will be considered in future publications.

\section{The RV measure for a landscape\label{sub:Extending-the-RV}}

The RV measure prescription as formulated in Refs.~\cite{Winitzki:2008yb,Winitzki:2008ph}
applies only to the calculation of abundances of terminal bubbles.
We will now extend the RV measure to computing arbitrary statistics
on a landscape.

We first note that RV measure prescription can be applied, strictly
speaking, only to landscapes that contain \emph{some} terminal bubble
types (i.e.~vacua from which no further tunneling is possible). However,
this limitation is quite benign, for two reasons. First, a landscape
without any Minkowski or AdS states is not expected to be realized
in any realistic string theory scenario without an exceptional amount
of fine-tuning. Second, the previously proposed volume-based and the
world-line based measure prescriptions agree for a landscape without
terminal bubble types~\cite{Vanchurin:2006qp,Bousso:2006ev}. One
may therefore consider the measure problem as solved in such landscapes
and turn one's attention to more realistic landscapes where terminal
bubble types are present.

Let us take an initial bubble of a nonterminal type $j$ and consider
the statistical ensemble $E_{n}(j)$ of all possible evolutions of
the initial bubble such that the total number of nucleated bubbles
of all types is finite and equals $n$ (not counting the initial bubble).
The total number of nucleated bubbles in a multiverse can be finite
only if terminal bubbles nucleate everywhere and merge globally to
the future of the initial bubble. This can happen by rare chance;
however, it is important the total probability of all events in the
ensemble $E_{n}(j)$ is always nonzero for any given $n$, so that
the ensembles $E_{n}(j)$ are well-defined and nonempty. 

The ensemble $E_{n}(j)$ may be described in the language of {}``transition
trees'' used in Ref.~\cite{Bousso:2006ev}. The ensemble consists
of all trees that have a total number $n+1$ of bubbles, including
the initial bubble of type $j$. The trees in $E_{n}(j)$ are finite
because all the {}``outer'' leaves are bubbles of terminal types.
The motivation for considering the ensemble $E_{n}(j)$ is that a
finite but very large tree (with $n$ large) is a controlled approximation
to infinite trees that typically occur. Hence, we are motivated to
consider $E_{n}(j)$ with $n$ finite but very large.

Note that the ensemble $E_{n}(j)$ differs from the ensemble defined
in Ref.~\cite{Winitzki:2008yb}; in $E_{n}(j)$ the total number
of bubbles of \emph{all} types is equal to $n$, rather than the total
number of \emph{terminal} bubbles as in Ref.~\cite{Winitzki:2008yb}.
Thus, the current proposal, which appears more natural, is an extension
of that of Ref.~\cite{Winitzki:2008yb}. Future work will show whether
this technical difference is significant; presently I will investigate
the consequences of the current proposal.

Once the ensemble $E_{n}(j)$ is defined, one may consider the statistical
distribution of some cosmological observable within the multiverses
belonging to the set $E_{n}(j)$. For instance, one can count the
number of bubbles of some type $k$, or the number of observers within
bubbles of type $k$, or the number of observations of some physical
process, etc. In a very large multiverse belonging to $E_{n}(j)$
with $n\gg1$, one may expect that the statistics of observations
will be independent of the initial bubble type $j$. Indeed, this
will be one of the results of this paper. Hence, let us suppress the
argument $j$ and write simply $E_{n}$.

Each multiverse belonging to $E_{n}$ has a naturally defined probability
weight, which is simply equal to the probability of realizing that
multiverse. This probability weight needs to be taken into account
when computing the statistical distributions of observables. The sum
of all probability weights of multiverses within $E_{n}$ is equal
to the total probability of $E_{n}$, which is exponentially small
for large $n$ but always nonzero. Since the multiverses from the
ensemble $E_{n}$ are by construction finite, i.e.~each multiverse
supports only a finite total number of possible observers, we are
assured that any statistics we desire to compute on $E_{n}$ will
be well-defined. 

We can now consider the probability distribution $p(Q|E_{n})$ of
some interesting observable $Q$ within the ensemble $E_{n}$ and
take the limit $n\rightarrow\infty$. One expects that the probability
distribution $p(Q|E_{n})$ will have a well-defined limit for large
$n$, \begin{equation}
p(Q)\equiv\lim_{n\rightarrow\infty}p(Q|E_{n}).\label{eq:PQ lim}\end{equation}
It was shown in previous work on the RV prescription~\cite{Winitzki:2008yb,Winitzki:2008ph}
that the distribution $p(Q)$ is well-defined for a simplest toy landscape
as well as in the case of random-walk eternal inflation. In this work
I extend these results to a general landscape scenario. Below (Sec.~\ref{sub:Uniqueness-of-the})
I will prove rigorously that the limit~(\ref{eq:PQ lim}) indeed
exists and is independent of the chosen initial bubble type $j$ as
long as the initial bubble is not of terminal type and as long as
the landscape is irreducible (every vacuum can be reached from every
other non-terminal vacuum by a chain of nucleations). Thus, the distribution
$p(Q)$ is unique and well-defined. This distribution is the final
result of applying the RV prescription to the observable $Q$. 

In practice, it is necessary to compute the distribution $p(Q|E_{n})$
asymptotically in the limit of large $n$. A direct numerical calculation
of probabilities in that limit by enumerating all possible evolution
trees is extremely difficult because of the exponential growth of
the number of possible evolutions. Instead, I derive explicit formulas
for the distribution $p(Q)$ in a generic landscape by evaluating
the limit~(\ref{eq:PQ lim}) analytically. These formulas are the
main result of the present article.

\section{A toy landscape\label{sub:A-toy-landscape}}

I begin by applying the RV measure to a toy landscape with very few
vacua. Using this simple example, I develop the computational techniques
needed for the practical evaluation of the limit such as Eq.~(\ref{eq:PQ lim}).
In Sec.~\ref{sec:A-general-landscape} the same techniques will be
extended to a more general landscape with an arbitrary number of vacua. 

In Ref.~\cite{Winitzki:2008yb} I considered the simplest possible
nontrivial landscape: a single dS and two AdS (terminal) vacua. The
next least complicated example that can be treated analytically is
a toy landscape having two nonterminal and two terminal vacua. This
toy landscape was called the \emph{{}``}FABI'' model in Ref.~\cite{Garriga:2005av}
and consists of the vacua labeled $F,I,A,B$ with the transition diagram
$A\leftarrow F\leftrightarrow I\rightarrow B$. In other words, one
assumes that the vacuum $F$ ({}``false'' vacuum) can nucleate only
bubbles of types $A$ and $I$, the vacuum $I$ ({}``intermediate''
vacuum) can nucleate bubbles of types $F$ and $B$, while $A$ and
$B$ are terminal vacua that do not have further nucleations. 

\begin{figure}
\begin{centering}\includegraphics[width=3.5in]{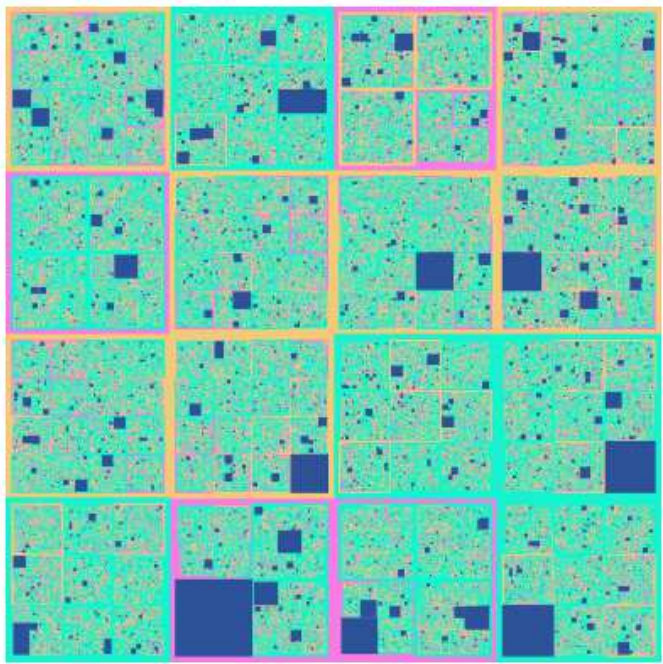}\par\end{centering}

\caption{An example simulation of {}``eternal inflation in a box'' in two
spatial dimensions. Bubbles ($H$-regions) are represented in comoving
coordinates by squares. Dark shades indicate bubbles of terminal types.
Other shades correspond to nested bubbles of various nonterminal ({}``recyclable'')
types. For the purposes of visual illustration, nucleation rates were
chosen of order one, and colored lines were drawn at bubble boundaries.
\label{fig:An-example-inflinabox}}
\end{figure}

To describe the finitely produced probability in this landscape, I
use the discrete picture called the {}``eternal inflation in a box''~\cite{Winitzki:2005ya,Winitzki:2008yb},
which is closely related to the {}``square bubble'' approximation~\cite{Bousso:2007nd}.
In this picture, one considers the evolution of discrete, causally
disjoint homogeneous $H$-regions in discrete time. All possible vacuum
types are labeled by $j=1,...,N$. During one time step of order $\delta t=H_{j}^{-1}$,
where $H_{j}$ is the local Hubble rate in a given $H$-region of
type $j$, the evolution consists of expanding the $H$-region into
\begin{equation}
e^{3H_{j}\delta t}=e^{3}\equiv\nu\end{equation}
 daughter $H$-regions of type $j$. Each of the daughter $H$-regions
has then the probability $\kappa_{j\rightarrow k}$ of changing immediately
into an $H$-region of type $k\neq j$; this imitates a nucleation
of a horizon-size bubble of type $k$. If no transition has taken
place, the daughter $H$-region retains its type $j$. For convenience,
we denote by \begin{equation}
\kappa_{j\rightarrow j}\equiv1-\sum_{k\neq j}\kappa_{j\rightarrow k}\label{eq:kappa jj def}\end{equation}
 the probability of no transitions during one Hubble time. The process
of expansion and transition is continued \emph{ad infinitum}, independently
for each resulting $H$-region. The $H$-regions of terminal types
will admit no further transitions and will not expand further (except,
perhaps, by a fixed amount due to slow-roll inflation occuring immediately
after nucleation). An example simulation is shown in Fig.~\ref{fig:An-example-inflinabox}.

In the remainder of this section I perform explicit calculations of
the RV cutoff in the FABI model.

\subsection{Bubble abundances\label{sub:Bubble-abundance}}

The first task is to compute the relative abundance of bubbles of
different types. Consider the probability $p(n_{\text{tot}},n_{F},n_{I},n_{A},n_{B};k)$
of having a finite total number $n_{\text{tot}}$ of bubbles of which
$n_{j}$ are of type $j$ (where $j=F,I,A,B$), if one starts from
a single initial $H$-region of type $k$ ($k=F,I$). By construction,
$p\neq0$ only for $n_{\text{tot}}=n_{F}+n_{I}+n_{A}+n_{B}$. A generating
function for this probability distribution can be defined by \begin{align}
g(z,\left\{ q_{j}\right\} ;k) & \equiv\negmedspace\sum_{n{\geq0,n}_{j}\geq0}\negmedspace z^{n}p(n,\left\{ n_{j}\right\} ;k)\negmedspace\prod_{j=F,I,A,B}\negmedspace q_{j}^{n_{j}}\nonumber \\
 & \equiv\left\langle z^{n}q_{F}^{n_{F}}q_{I}^{n_{I}}q_{A}^{n_{A}}q_{B}^{n_{B}}\right\rangle _{n<\infty;k},\label{eq:g FABI def}\end{align}
where the notation $\left\langle ...\right\rangle _{n<\infty;k}$
stands for a statistical average restricted to events with a finite
total number $n$ of bubbles nucleated to the future of an initial
bubble of type $k$.

The generating function $g$ plays a crucial role in the entire calculation.
Since we will be only interested in the initial bubbles of types $F$
and $I$, let us denote \begin{equation}
F(z,\left\{ q_{j}\right\} )\equiv g(z,\left\{ q_{j}\right\} ;F),\quad I(z,\left\{ q_{j}\right\} )\equiv g(z,\left\{ q_{j}\right\} ;I).\end{equation}
The generating functions $F$ and $I$ satisfy the following system
of nonlinear algebraic equations~\cite{Winitzki:2008yb},\begin{align}
F^{\frac{1}{\nu}} & =zq_{A}\kappa_{FA}+zq_{I}\kappa_{FI}I+\kappa_{FF}F,\label{eq:F equ 1}\\
I^{\frac{1}{\nu}} & =zq_{B}\kappa_{IB}+zq_{F}\kappa_{IF}F+\kappa_{II}I.\label{eq:I equ 1}\end{align}
Here we denoted for brevity $\kappa_{FF}\equiv1-\kappa_{FA}-\kappa_{FI}$
and $\kappa_{II}\equiv1-\kappa_{IB}-\kappa_{IF}$; within our assumptions,
$\kappa_{FF}\approx1$ and $\kappa_{II}\approx1$. In Eqs.~(\ref{eq:F equ 1})--(\ref{eq:I equ 1})
the generating variable $z$ multiplies only the terms that correspond
to changing the type of the $H$-region (which imitates the nucleation
of new bubbles) but not the terms $\kappa_{FF}F$ and $\kappa_{II}I$
that correspond to the eventuality of not changing the type of the
$H$-region during one Hubble time.

The nonlinear equations~(\ref{eq:F equ 1})--(\ref{eq:I equ 1})
may have several real-valued solutions, as well as complex-valued
solutions that are certainly not of physical interest. In particular,
for $z=1$ and $q_{j}=1$ there exists the {}``trivial'' solution
$F=I=1$ as well as a nontrivial solution with $F\ll1$ and $I\ll1$.
Similarly, for $z$ near 0 there exists the solution that approaches
$F(0)=I(0)=0$,\begin{equation}
F=\left(zq_{A}\kappa_{FA}\right)^{\nu}+O(z^{2\nu-1}),\; I\approx\left(zq_{B}\kappa_{IB}\right)^{\nu}+O(z^{2\nu-1}),\label{eq:approx sol F I at z0}\end{equation}
as well as the solution \begin{equation}
F\approx\kappa_{FF}^{\frac{\nu}{1-\nu}}\approx1,\quad I\approx\kappa_{II}^{\frac{\nu}{1-\nu}}\approx1\label{eq:approx sol bad}\end{equation}
and solutions where $F\approx1$ and $I\approx0$ and vice versa.
It is important to determine the solution branch $F(z),I(z)$ that
has the physical significance as the actual generating function of
the finitely produced distribution of $H$-regions. 

The functions $F(z),I(z)$ are solutions of algebraic equations and
thus are continuous functions of $z$ that are analytic everywhere
in complex $z$ plane except for branch cuts. It is easy to see that
the solution $F=I=1$ at $z=1,q_{j}=1$ is continuously connected
with the solution~(\ref{eq:approx sol bad}) at $z\approx0$, while
the solution~(\ref{eq:approx sol F I at z0}) is continued to a solution
with $F(z)\ll1$ and $I(z)\ll1$ for all $0<z<1$. The values $F(1)$
and $I(1)$ are the probabilities of the events that the evolution
of an initial bubble of type $F$ or $I$ ends globally. These probabilities
are extremely small and of order $\kappa_{FA}^{\nu}$ and $\kappa_{IB}^{\nu}$
respectively. This is easy to interpret because, for instance, $\kappa_{FA}^{\nu}$
is the probability of nucleating $\nu$ terminal regions at once after
one Hubble time within an $H$-region of type $F$. Hence, the generating
functions for the finitely produced distribution of bubbles are given
by the solution branch having $F\ll1$ and $I\ll1$ rather than by
the solution $F=I=1$ at $z=1$. More precisely, the physically meaningful
solution $F(z),I(z)$ is selected by the asymptotic behavior~(\ref{eq:approx sol F I at z0}).
This argument removes the ambiguity inherent in solving Eqs.~(\ref{eq:F equ 1})--(\ref{eq:I equ 1}).
Below we refer to the branch of solutions $F(z),I(z)$ connected to
the nontrivial solution~(\ref{eq:approx sol F I at z0}) near $z=0$
as the {}``main branch.''

Once the main branch of the generating functions $g(z,\{ q_{j}\};k)$
are known, one can express the mean number of bubbles of type $i$
($i=F,I,A,B$) at a fixed total number $n_{\text{tot}}$ as \begin{equation}
p(i|n_{\text{tot}})\equiv\frac{\left\langle n_{i}\right\rangle _{n_{\text{tot}}}}{n_{\text{tot}}}=\frac{\partial_{z}^{n_{\text{tot}}}\partial_{q_{i}}g(z=0,\left\{ q_{j}=1\right\} ;k)}{n_{\text{tot}}\partial_{z}^{n_{\text{tot}}}g(z=0,\left\{ q_{j}=1\right\} ;k)}.\label{eq:p n' n lim}\end{equation}
One expects that the limit of this ratio at $n_{\text{tot}}\rightarrow\infty$
will be independent of the initial bubble type $k$ since the ensemble
$E_{n_{\text{tot}}}$ will consist of $H$-regions having a very long
evolution, so that the initial conditions are forgotten. Below I will
show explicitly that this is indeed the case.

It is more convenient to compute the ratios of the number of bubbles
of types $i$ and $i'$ at fixed $n_{\text{tot}}$,\begin{equation}
\frac{p(i|n_{\text{tot}})}{p(i'|n_{\text{tot}})}=\left.\frac{\partial_{z}^{n_{\text{tot}}}\partial_{q_{i}}g}{\partial_{z}^{n_{\text{tot}}}\partial_{q_{i'}}g}\right|_{z=0,\left\{ q_{j}=1\right\} }.\label{eq:i i' ratio ntot}\end{equation}
Then one only needs to compute derivatives $\partial_{q_{i}}g\equiv g_{,q_{i}}$
evaluated at $q_{j}=1$. These derivatives satisfy a system of \emph{linear}
equations that can be easily derived from Eqs.~(\ref{eq:F equ 1})--(\ref{eq:I equ 1}).
For instance, the derivatives $F_{,q_{A}}$ and $I_{,q_{A}}$ satisfy\begin{align}
\frac{1}{\nu}F^{\frac{1}{\nu}-1}F_{,q_{A}} & =z\kappa_{FA}+z\kappa_{FI}I_{,q_{A}}+\kappa_{FF}F_{,q_{A}},\\
\frac{1}{\nu}I^{\frac{1}{\nu}-1}I_{,q_{A}} & =z\kappa_{IF}F_{,q_{A}}+\kappa_{II}I_{,q_{A}}.\end{align}
Rewriting these equations in a matrix form, we obtain\begin{equation}
\left(\begin{array}{cc}
\frac{1}{\nu}F^{\frac{1}{\nu}-1}-\kappa_{FF} & -z\kappa_{FI}\\
-z\kappa_{IF} & \frac{1}{\nu}I^{\frac{1}{\nu}-1}-\kappa_{II}\end{array}\right)\left[\begin{array}{c}
F_{,q_{A}}\\
I_{,q_{A}}\end{array}\right]=\left[\begin{array}{c}
z\kappa_{FA}\\
0\end{array}\right].\label{eq:FIqA eqs}\end{equation}
The coefficients of the $z$-dependent matrix\begin{equation}
\hat{M}(z)\equiv\left(\begin{array}{cc}
\frac{1}{\nu}F^{\frac{1}{\nu}-1}-\kappa_{FF} & -z\kappa_{FI}\\
-z\kappa_{IF} & \frac{1}{\nu}I^{\frac{1}{\nu}-1}-\kappa_{II}\end{array}\right)\end{equation}
 are the main branch of solutions of the nonlinear equations~(\ref{eq:F equ 1})--(\ref{eq:I equ 1})
at $q_{j}=1$ but at arbitrary $z$. 

One can verify using Eq.~(\ref{eq:approx sol F I at z0}) that the
matrix $\hat{M}(z)$ is invertible near $z=0$. Hence the solution
of Eq.~(\ref{eq:FIqA eqs}) can be written, at least within some
range of $z$ where $\hat{M}(z)$ remains invertible, as\begin{equation}
\left[\begin{array}{c}
F_{,q_{A}}\\
I_{,q_{A}}\end{array}\right]=\hat{M}^{-1}(z)\left[\begin{array}{c}
z\kappa_{FA}\\
0\end{array}\right].\label{eq:FqA eq 1}\end{equation}
Similarly, the derivatives $F_{,q_{F}}$ and $I_{,q_{F}}$ satisfy
the equations\begin{equation}
\hat{M}(z)\left[\begin{array}{c}
F_{,q_{F}}\\
I_{,q_{F}}\end{array}\right]=\left[\begin{array}{c}
0\\
z\kappa_{IF}F\end{array}\right],\label{eq:FqF equ 1 pre}\end{equation}
whose solution is\begin{equation}
\left[\begin{array}{c}
F_{,q_{F}}\\
I_{,q_{F}}\end{array}\right]=\hat{M}^{-1}(z)\left[\begin{array}{c}
0\\
z\kappa_{IF}F\end{array}\right].\label{eq:FqF eq 1}\end{equation}
Other generating functions can be expressed in the same manner.

One could in principle obtain a numerical solution for $F_{,q_{A}}(z)$,
$I_{,q_{A}}(z)$, and all the other generating functions at any given
value of $z$. However, the numerical solution is not particularly
useful at this point because the next step in the calculation is the
evaluation of Eq.~(\ref{eq:i i' ratio ntot}) in the limit of very
large $n_{\text{tot}}$, for instance,\begin{equation}
\frac{p(A|n_{\text{tot}})}{p(F|n_{\text{tot}})}=\left.\frac{\partial_{z}^{n_{\text{tot}}}F_{,q_{A}}}{\partial_{z}^{n_{\text{tot}}}F_{,q_{F}}}\right|_{z=0}.\label{eq:ratio a to f}\end{equation}
It is generally not feasible to compute the $n$-th derivative of
a numerically obtained function in the limit $n\rightarrow\infty$,
because the unavoidable round-off errors are amplified by a fixed
factor with each successive numerical differentiation. Therefore,
we need a way to evaluate the ratio~(\ref{eq:i i' ratio ntot}) in
the limit $n_{\text{tot}}\rightarrow\infty$ without using numerics.
Indeed we will be able to compute the ratio $p(A)/p(F)$. We will
also show that, for instance,\begin{equation}
\frac{p(A)}{p(F)}=\lim_{n\rightarrow\infty}\left.\frac{\partial_{z}^{n}F_{,q_{A}}}{\partial_{z}^{n}F_{,q_{F}}}\right|_{z=0}=\lim_{n\rightarrow\infty}\left.\frac{\partial_{z}^{n}I_{,q_{A}}}{\partial_{z}^{n}I_{,q_{F}}}\right|_{z=0};\label{eq:ratio a to f lim}\end{equation}
in other words, that the final RV-regulated ratio $p(A)/p(F)$ is
independent of whether the initial bubble is of type $F$ or of type
$I$.

To proceed, we use the fact that $F_{,q_{A}}(z)$, $I_{,q_{A}}(z)$,
$F_{,q_{F}}(z)$, etc.~are analytic functions of the parameter $z$.
The asymptotic growth of high-order derivatives of an analytic function
$f(z)$ is determined by the location of the singularities of $f(z)$
in the complex $z$ plane. The required result can be derived by the
following elementary argument: Consider the derivative $d^{n}f/dz^{n}$
at $z=0$, and assume that $f(z)$ admits an expansion around the
singularity $z_{*}$ nearest to $z=0$, such as \begin{equation}
f(z)=c_{0}+c_{1}\left(z-z_{*}\right)^{s}+...,\label{eq:f sing}\end{equation}
where $s\neq0,1,2,...$ is the power of the leading-order singularity,
and the omitted terms are either higher powers of $z-z_{*}$ or singularities
at points $z_{*}^{\prime}$ located further away from $z=0$. The
singularity structure~(\ref{eq:f sing}) yields the large-$n$ asymptotics\begin{equation}
\left.\frac{d^{n}f}{dz^{n}}\right|_{z=0}\negmedspace\approx c_{1}(-z_{*})^{s}\frac{\Gamma(n-s)}{\Gamma(-s)}z_{*}^{-n}+...\label{eq:large-n derivative}\end{equation}
It can be seen from this formula that any other singular point $z_{*}^{\prime}$
located further away from $z=0$ gives a contribution that is smaller
by the factor $\left|z_{*}^{\prime}/z_{*}\right|^{s-n}$. The contribution
of a subdominant singularity of the form $\left(z-z_{*}\right)^{s'}$,
i.e.~at the same point $z=z_{*}$ but with a higher power $s'>s$,
is suppressed, in comparison with the term in Eq.~(\ref{eq:large-n derivative}),
by the factor \begin{equation}
\frac{\Gamma(n-s')}{\Gamma(n-s)}\frac{\Gamma(-s)}{\Gamma(-s')}\approx\frac{1}{\left(n-1\right)^{s'-s}}\frac{\Gamma(-s)}{\Gamma(-s')}.\end{equation}
It is clear that the terms omitted from Eq.~(\ref{eq:large-n derivative})
indeed give subleading contributions at large $n$, and so Eq.~(\ref{eq:large-n derivative})
is indeed the leading asymptotic term at $n\rightarrow\infty$. 

We now need to determine the location of the singularities of $F_{,q_{A}}(z)$,
$I_{,q_{A}}(z)$, etc., as functions of $z$. This task is much simplified
once we observe that all these quantities are expressed through the
inverse matrix $\hat{M}^{-1}(z)$, and hence it remains to analyze
the singularities of $\hat{M}^{-1}(z)$. That matrix can be singular
at some value $z=z_{*}$ either because some of the coefficients of
$\hat{M}(z)$ are singular, or because the matrix $\hat{M}(z)$ is
degenerate (noninvertible) at $z=z_{*}$. The coefficients of $\hat{M}(z)$
depend on solutions $F(z),I(z)$ of algebraic equations~(\ref{eq:F equ 1})--(\ref{eq:I equ 1})
at $q_{j}=1$ and thus cannot be divergent as functions of $z$. These
coefficients can be singular only in that some derivative in $z$
diverges. The derivatives $F_{,z}$ and $I_{,z}$ satisfy the equations\begin{equation}
\hat{M}(z)\left[\begin{array}{c}
F_{,z}\\
I_{,z}\end{array}\right]=\left[\begin{array}{c}
\kappa_{FA}+\kappa_{FI}I\\
\kappa_{IB}+\kappa_{IF}F\end{array}\right].\label{eq:equations for FI derivatives}\end{equation}
Therefore, $F_{,z}$ and $I_{,z}$ diverge only for those $z$ for
which the matrix $\hat{M}(z)$ is degenerate. For these $z$, all
the derivatives $F_{,z}$, $I_{,z}$, $F_{,q_{A}}$, etc.~will be
divergent at the same time since they are all proportional to the
inverse matrix $\hat{M}^{-1}(z)$.

We note that the matrix $\hat{M}(z)$ is the Jacobian of the nonlinear
system~(\ref{eq:F equ 1})--(\ref{eq:I equ 1}). As long as $\hat{M}(z)$
is nondegenerate, all the different branches of the solutions $F(z),I(z)$
do not meet and remain smooth functions of $z$. As we noted above,
the {}``main'' branch $F(z),I(z)$ is the one connected to the nontrivial
solution~(\ref{eq:approx sol F I at z0}) at $z\approx0$. The matrix
$\hat{M}(z)$ is nondegenerate near $z=0$; therefore, the solutions
$F(z),I(z)$ remain smooth functions of $z$ for all $z$ such that
$\hat{M}(z)$ is nondenegerate. We conclude that the only possible
singularities of $F(z),I(z)$ are those values $z=z_{*}$ where $\det\hat{M}(z)=0$.

Below it will be shown that the behavior of $\det\hat{M}(z)$ near
$z=z_{*}$ is \begin{equation}
\det\hat{M}(z)\approx c_{1}\sqrt{z_{*}-z},\label{eq:det M sing 1}\end{equation}
i.e.~of the form~(\ref{eq:f sing}) with $c_{0}=0$ and $s=\frac{1}{2}$.
It now follows from Eqs.~(\ref{eq:FqA eq 1}) and (\ref{eq:FqF eq 1})
that the derivatives such as $F_{,q_{A}}(z)$, $I_{,q_{A}}(z)$, $F_{,q_{F}}(z)$,
etc., all diverge at $z=z_{*}$ with the same asymptotic behavior,
namely proportional to $\left(z_{*}-z\right)^{-1/2}$. We may express
the inverse matrix as\begin{align}
\hat{M}^{-1}(z) & =\frac{1}{\det\hat{M}(z)}\hat{\tilde{M}}(z),\\
\hat{\tilde{M}}(z) & \equiv\left(\begin{array}{cc}
\frac{1}{\nu}I^{\frac{1}{\nu}-1}-\kappa_{II} & z\kappa_{FI}\\
z\kappa_{IF} & \frac{1}{\nu}F^{\frac{1}{\nu}-1}-\kappa_{FF}\end{array}\right),\label{eq:cofactor matrix}\end{align}
where we introduced the algebraic cofactor matrix $\hat{\tilde{M}}\equiv\hat{M}^{-1}\det\hat{M}$,
which is nonsingular at $z=z_{*}$. It then follows from Eq.~(\ref{eq:FqA eq 1})
that the asymptotic behavior of the functions $F_{,q_{A}}(z)$ and
$I_{,q_{A}}(z)$ near $z=z_{*}$ is given by \begin{align}
\left[\begin{array}{c}
F_{,q_{A}}\\
I_{,q_{A}}\end{array}\right] & \approx\frac{1}{c_{1}\sqrt{z_{*}-z}}\hat{\tilde{M}}(z_{*})\left[\begin{array}{c}
z_{*}\kappa_{FA}\\
0\end{array}\right]\nonumber \\
 & =\frac{1}{c_{1}\sqrt{z_{*}-z}}\left[\begin{array}{c}
\left(\frac{1}{\nu}I^{\frac{1}{\nu}-1}-\kappa_{II}\right)z_{*}\kappa_{FA}\\
z_{*}^{2}\kappa_{IF}\kappa_{FA}\end{array}\right].\end{align}
Similarly, using Eq.~(\ref{eq:FqF eq 1}) we find near $z=z_{*}$\begin{align}
\left[\begin{array}{c}
F_{,q_{F}}\\
I_{,q_{F}}\end{array}\right] & \approx\frac{1}{c_{1}\sqrt{z_{*}-z}}\hat{\tilde{M}}(z_{*})\left[\begin{array}{c}
0\\
z_{*}\kappa_{IF}F\end{array}\right]\nonumber \\
 & =\frac{1}{c_{1}\sqrt{z_{*}-z}}\left[\begin{array}{c}
z_{*}^{2}\kappa_{FI}\kappa_{IF}F\\
\left(\frac{1}{\nu}F^{\frac{1}{\nu}-1}-\kappa_{FF}\right)z_{*}\kappa_{IF}F\end{array}\right].\end{align}
 Using Eq.~(\ref{eq:large-n derivative}) and noticing that all $n$-dependent
factors are the same in $\partial_{z}^{n}F_{,q_{A}}$ and other such
derivatives, we conclude that \begin{align}
\lim_{n\rightarrow\infty}\left.\frac{\partial_{z}^{n}F_{,q_{A}}}{\partial_{z}^{n}F_{,q_{F}}}\right|_{z=0} & =F^{\frac{1}{\nu}}\frac{\left(\frac{1}{\nu}I_{*}^{\frac{1}{\nu}-1}-\kappa_{II}\right)\kappa_{FA}}{z_{*}\kappa_{FI}\kappa_{IF}F_{*}},\\
\lim_{n\rightarrow\infty}\left.\frac{\partial_{z}^{n}I_{,q_{A}}}{\partial_{z}^{n}I_{,q_{F}}}\right|_{z=0} & =\frac{z_{*}\kappa_{FA}}{\left(\frac{1}{\nu}F_{*}^{\frac{1}{\nu}-1}-\kappa_{FF}\right)F_{*}}.\end{align}
It is important that the two limits above are equal; this is so because
the condition $\det\hat{M}(z_{*})=0$ yields \begin{equation}
\det\hat{M}=\left[\frac{F_{*}^{\frac{1}{\nu}-1}}{\nu}-\kappa_{FF}\right]\negmedspace\left[\frac{I_{*}^{\frac{1}{\nu}-1}}{\nu}-\kappa_{II}\right]-z_{*}^{2}\kappa_{FI}\kappa_{IF}=0,\label{eq:det M equ 0}\end{equation}
where we denoted $F_{*}\equiv F(z_{*})$, $I_{*}\equiv I(z_{*})$
for brevity. It follows that the ratio of $A$-bubbles to $F$-bubbles
is independent of whether the initial bubble is of type $F$ or of
type $I$. In other words, the RV-regulated ratio of $A$-bubbles
to $F$-bubbles is\begin{equation}
\frac{p(A)}{p(F)}=\frac{z_{*}\kappa_{FA}}{\left(\frac{1}{\nu}F_{*}^{\frac{1}{\nu}-1}-\kappa_{FF}\right)F_{*}}\label{eq:AF ratio}\end{equation}
independently of the initial bubble type. Below (Sec.~\ref{sub:Uniqueness-of-the})
the independence of initial conditions will be rigorously proved for
a general landscape using mathematical techniques of the theory of
nonnegative matrices. Presently we have shown this independence using
explicit formulas available for the toy landscape under consideration. 

In a similar way, we find the RV-regulated ratio of $A$-bubbles to
$B$-bubbles,\begin{equation}
\frac{p(A)}{p(B)}=\frac{z_{*}\kappa_{FA}\kappa_{IF}}{\left(\frac{1}{\nu}F_{*}^{\frac{1}{\nu}-1}-\kappa_{FF}\right)\kappa_{IB}},\label{eq:AB ratio}\end{equation}
and the ratio of $F$-bubbles to $I$-bubbles,\begin{equation}
\frac{p(F)}{p(I)}=\frac{\left(\frac{1}{\nu}F_{*}^{\frac{1}{\nu}-1}-\kappa_{FF}\right)F_{*}}{z_{*}\kappa_{FI}I_{*}}.\label{eq:FI ratio}\end{equation}

It remains to compute $z_{*}$, $F_{*}$, $I_{*}$ and to justify
Eq.~(\ref{eq:det M sing 1}). We will do this using the explicit
form of the matrix $\hat{M}(z)$. To determine $z_{*},F_{*}$, and
$I_{*}$, we need to solve Eq.~(\ref{eq:det M equ 0}) simultaneously
with the equations \begin{align}
F_{*}^{\frac{1}{\nu}} & =z_{*}\kappa_{FA}+z_{*}\kappa_{FI}I_{*}+\kappa_{FF}F_{*},\label{eq:F equ 2}\\
I_{*}^{\frac{1}{\nu}} & =z_{*}\kappa_{IB}+z_{*}\kappa_{IF}F_{*}+\kappa_{II}I_{*},\label{eq:I equ 2}\end{align}
the latter two being Eqs.~(\ref{eq:F equ 1})--(\ref{eq:I equ 1})
after setting $q_{j}=1$. We are interested in the solution $z_{*}$
closest to $z=0$. By definition~(\ref{eq:g FABI def}), the generating
functions are nonsingular for $\left|z\right|\leq1$; hence, the only
possible values of $z_{*}$ are in the domain $\left|z\right|>1$
in the complex plane.

Moreover, we can show that the solutions $F(z),I(z)$ are growing
functions of $z$ for real $z<z_{*}$. Initially at $z\approx0$ these
functions have the form~(\ref{eq:approx sol F I at z0}) and hence
are positive. The determinant $\det\hat{M}$ shown in Eq.~(\ref{eq:det M equ 0})
is also positive for $z<z_{*}$. One can then find from Eq.~(\ref{eq:cofactor matrix})
that the algebraic cofactor matrix $\hat{\tilde{M}}$ and hence the
inverse matrix $\hat{M}^{-1}$ has all positive elements for those
$z$. It is then evident from Eq.~(\ref{eq:equations for FI derivatives})
that $\partial_{z}F>0$ and $\partial_{z}I>0$ as long as the values
of $F$ and $I$ are themselves positive. Therefore, the solutions
$F(z),I(z)$ are positive and growing functions of $z$ as long as
$\det\hat{M}(z)>0$. 

We may thus visualize the behavior of $\det\hat{M}(z)$ as $z$ grows.
Both $F(z)$ and $I(z)$ will grow with $z$, so that the terms in
square brackets diminish while the second term, $z_{*}^{2}\kappa_{FI}\kappa_{IF}$,
grows. Eventually the product of the square brackets in Eq.~(\ref{eq:det M equ 0})
will be balanced by the second term, and the determinant will vanish.
We need to determine the smallest value $z=z_{*}$ for which $\det\hat{M}(z)=0$.

Since the physically significant branch of the solution involves always
very small values $F(z)$ and $I(z)$ for $\left|z\right|\leq1$,
while other (unphysical) branches have either $F$ or $I$ approximately
equal to 1, it is reasonable to assume that $F(z_{*})\ll1$ and $I(z_{*})\ll1$
also at $z=z_{*}$ (the self-consistency of this assumption will be
confirmed by later calculations). Then the terms $\kappa_{FF}F_{*}\approx F_{*}$
and $\kappa_{II}I_{*}\approx I_{*}$ can be disregarded in comparison
with $F_{*}^{1/\nu}$ and $I_{*}^{1/\nu}$ in Eqs.~(\ref{eq:F equ 2})--(\ref{eq:I equ 2}).
In this approximation, we can simplify Eqs.~(\ref{eq:F equ 2})--(\ref{eq:I equ 2})
to\begin{align}
F_{*}^{\frac{1}{\nu}} & =z_{*}\kappa_{FA}+z_{*}\kappa_{FI}I_{*},\label{eq:F equ 2 s}\\
I_{*}^{\frac{1}{\nu}} & =z_{*}\kappa_{IB}+z_{*}\kappa_{IF}F_{*},\label{eq:I equ 2 s}\end{align}
 while Eq.~(\ref{eq:det M equ 0}) becomes \begin{equation}
\left(\kappa_{FA}+\kappa_{FI}I_{*}\right)\left(\kappa_{IB}+\kappa_{IF}F_{*}\right)=\nu^{2}\kappa_{FI}\kappa_{IF}F_{*}I_{*}.\label{eq:detM 0 simp}\end{equation}
The ratios~(\ref{eq:AF ratio})--(\ref{eq:FI ratio}) are also simplified
and can be written more concisely as\begin{align}
p(A):p(I) & :p(F):p(B)=\kappa_{FA}:\left(\kappa_{FI}I_{*}\right)\nonumber \\
 & :\frac{\kappa_{FA}+\kappa_{FI}I_{*}}{\nu}:\kappa_{IB}\frac{\kappa_{FA}+\kappa_{FI}I_{*}}{\nu\kappa_{IF}F_{*}}.\label{eq:combined ratios}\end{align}
To determine $z_{*}$, we will obtain an explicit approximation for
the main branch $F(z),I(z)$ for all $z$. For small enough $z$,
the solutions of Eqs.~(\ref{eq:F equ 2})--(\ref{eq:I equ 2}) are
well approximated by Eq.~(\ref{eq:approx sol F I at z0}), \begin{equation}
F(z)\approx\left(z\kappa_{FA}\right)^{\nu},\quad I(z)\approx\left(z\kappa_{IB}\right)^{\nu}.\label{eq:regime 1 sol}\end{equation}
These solutions are obtained under the assumption that the terms $z\kappa_{FA}$
and $z\kappa_{IB}$ are numerically small ($z\kappa_{FA}\ll1$, $z\kappa_{IB}\ll1$)
and yet dominant in Eqs.~(\ref{eq:F equ 2})--(\ref{eq:I equ 2}).
These terms only remain dominant as long as \begin{equation}
\kappa_{FA}\gg\kappa_{FI}I(z),\quad\kappa_{IB}\gg\kappa_{IF}F(z).\label{eq:regime 1}\end{equation}
Substituting Eq.~(\ref{eq:approx sol F I at z0}) for $F(z)$ and
$I(z)$ into Eq.~(\ref{eq:regime 1}), we obtain the conditions\begin{equation}
z\ll\kappa_{IB}^{-1}\left(\frac{\kappa_{FA}}{\kappa_{FI}}\right)^{\frac{1}{\nu}},\quad z\ll\kappa_{FA}^{-1}\left(\frac{\kappa_{IB}}{\kappa_{IF}}\right)^{\frac{1}{\nu}}.\label{eq:regime 1 cond}\end{equation}
We need to check whether $\det\hat{M}(z)$ could vanish already for
some $z_{*}$ within the range~(\ref{eq:regime 1 cond}). Using Eq.~(\ref{eq:detM 0 simp})
in the regime~(\ref{eq:regime 1}), we find \begin{equation}
z_{*}\approx\left[\left(\kappa_{FA}\kappa_{IB}\right)^{\nu-1}\nu^{2}\kappa_{FI}\kappa_{IF}\right]^{-\frac{1}{2\nu}}.\end{equation}
This value is within the range~(\ref{eq:regime 1 cond}) only if
the following simultaneous inequalities hold, \begin{equation}
e^{-6}\equiv\frac{1}{\nu^{2}}\ll\left(\frac{\kappa_{IB}}{\kappa_{FA}}\right)^{\nu+1}\frac{\kappa_{FI}}{\kappa_{IF}}\ll\nu^{2}\equiv e^{6}.\label{eq:regime 1 cond 1}\end{equation}
Let us denote by $\eta$ the quantity in Eq.~(\ref{eq:regime 1 cond 1}),\begin{equation}
\eta\equiv\left(\frac{\kappa_{IB}}{\kappa_{FA}}\right)^{\nu+1}\frac{\kappa_{FI}}{\kappa_{IF}},\end{equation}
then Eq.~(\ref{eq:regime 1 cond 1}) becomes simply $\left|\ln\eta\right|<6$.
One would expect that $\eta$ is generically either very large or
very small, so the inequalities~(\ref{eq:regime 1 cond 1}) can hold
only in a fine-tuned landscape. Additionally, we need to require \begin{equation}
z_{*}\ll\min\left(\frac{1}{\kappa_{FA}},\frac{1}{\kappa_{IB}}\right),\end{equation}
which entails\begin{equation}
\frac{\kappa_{IB}}{\kappa_{IF}}\ll\nu\sqrt{\eta},\quad\frac{\kappa_{FA}}{\kappa_{FI}}\ll\frac{\nu}{\sqrt{\eta}}.\label{eq:regime 1 extra}\end{equation}
However, this requirement is weaker than Eq.~(\ref{eq:regime 1 cond 1})
since typically $\kappa_{FA}\ll\kappa_{FI}$ and $\kappa_{IB}\ll\kappa_{IF}$,
so\begin{equation}
z_{*}\ll\kappa_{IB}^{-1}\left(\frac{\kappa_{FA}}{\kappa_{FI}}\right)^{\frac{1}{\nu}}\ll\kappa_{IB}^{-1};\: z\ll\kappa_{FA}^{-1}\left(\frac{\kappa_{IB}}{\kappa_{IF}}\right)^{\frac{1}{\nu}}\ll\kappa_{FA}^{-1}.\end{equation}
 Let us assume, for the moment, that Eqs.~(\ref{eq:regime 1 cond 1})
and~(\ref{eq:regime 1 extra}) hold. Then we use Eq.~(\ref{eq:combined ratios})
to obtain \begin{equation}
p(A):p(I):p(F):p(B)\approx1:\frac{\sqrt{\eta}}{\nu}:\frac{1}{\nu}:\sqrt{\eta}.\label{eq:regime1 ans fine-tuned}\end{equation}
Due to the fine-tuning assumption~(\ref{eq:regime 1 cond 1}), these
ratios are all within the interval $\left[\nu^{-2},\nu^{2}\right]=\left[e^{-6},e^{6}\right]$. 

Having considered the fine-tuned case, let us now turn to the generic
case. Generically one would expect that the quantity $\eta$ is either
extremely large or extremely small, and in any case outside the logarithmically
narrow range $[e^{-6},e^{6}]$. In that case, one of the inequalities~(\ref{eq:regime 1 cond 1})
does not hold; generically, either $\eta\ll\nu^{-2}$ or $\eta\gg\nu^{2}$.
It follows that $\det\hat{M}(z)\neq0$ for all $z$ within the range~(\ref{eq:regime 1 cond}).
To reach the value $z_{*}$ at which $\det\hat{M}(z_{*})=0$, we need
to increase $z$ further, until one of the terms $z\kappa_{FI}I$
or $z\kappa_{IF}F$ in Eqs.~(\ref{eq:F equ 2})--(\ref{eq:I equ 2})
becomes dominant and the solution~(\ref{eq:regime 1 sol}) becomes
invalid.

It is impossible that \emph{both} the terms $z\kappa_{FI}I$ and $z\kappa_{IF}F$
are dominant in Eqs.~(\ref{eq:F equ 2})--(\ref{eq:I equ 2}) at
$z=z_{*}$ because then Eq.~(\ref{eq:detM 0 simp}) would yield a
contradiction,\begin{equation}
\kappa_{FI}I_{*}\kappa_{IF}F_{*}=\nu^{2}\kappa_{FI}\kappa_{IF}F_{*}I_{*}.\end{equation}
Hence, the determinant $\det\hat{M}(z)$ first vanishes at the value
$z=z_{*}$ such that only \emph{one} of those terms is dominant in
its respective equation. Without loss of generality, we may relabel
the vacua ($F\leftrightarrow I$, $A\leftrightarrow B$) such that
the term $z\kappa_{IF}F$ becomes dominant in Eq.~(\ref{eq:I equ 2})
while the term $z\kappa_{FA}$ is still dominant in Eq.~(\ref{eq:F equ 2}).
This is equivalent to assuming $\eta\ll\nu^{-2}$. In the range of
$z$ for which this is the case, \begin{equation}
z\kappa_{FI}I(z)\ll z\kappa_{FA}\ll1;\quad z\kappa_{IB}\ll z\kappa_{IF}F(z)\ll1,\label{eq:regime 2}\end{equation}
the approximate solution of Eqs.~(\ref{eq:F equ 2})--(\ref{eq:I equ 2})
can be written as\begin{equation}
F(z)\approx\left(z\kappa_{FA}\right)^{\nu};\; I(z)\approx\left(z\kappa_{IF}F\right)^{\nu}=\kappa_{FA}^{\nu^{2}}\kappa_{IF}^{\nu}z^{\left(\nu+1\right)\nu}.\label{eq:regime 2 sol}\end{equation}
With these values of $F(z)$ and $I(z)$, the consistency requirement~(\ref{eq:regime 2})
yields the following range of $z$,\begin{equation}
\frac{\kappa_{IB}}{\kappa_{IF}\kappa_{FA}^{\nu}}\ll z^{\nu}\ll\left[\kappa_{FA}^{\nu^{2}-1}\kappa_{FI}\kappa_{IF}^{\nu}\right]^{-\frac{1}{\nu+1}}.\label{eq:regime 2 cond}\end{equation}
This range is nonempty if \begin{equation}
\eta\equiv\left(\frac{\kappa_{IB}}{\kappa_{FA}}\right)^{\nu+1}\frac{\kappa_{FI}}{\kappa_{IF}}\ll1,\end{equation}
which is indeed one of the two possible ways that the fine-tuning~(\ref{eq:regime 1 cond 1})
can fail. Having assumed that the above condition holds, we need to
determine the value of $z_{*}$ and check that it belongs to the range~(\ref{eq:regime 2}).
Using Eq.~(\ref{eq:detM 0 simp}) in the regime~(\ref{eq:regime 2}),
we obtain\begin{equation}
z_{*}\approx\left[\nu^{2}\kappa_{IF}^{\nu}\kappa_{FI}\kappa_{FA}^{\nu^{2}-1}\right]^{-\frac{1}{\nu\left(\nu+1\right)}}.\end{equation}
Substituting this value into the inequalities~(\ref{eq:regime 2 cond}),
we find the condition \begin{equation}
\eta\ll\frac{1}{\nu^{2}}\ll1,\end{equation}
which holds identically under the current assumption, $\eta\ll\nu^{-2}$.
(The relabeling $F\leftrightarrow I$, $A\leftrightarrow B$ is necessary
if the opposite case, $\eta\gg\nu^{2}$, holds.) Therefore, $z_{*}$
is within the regime~(\ref{eq:regime 2}), and our approximations
are self-consistent, yielding \begin{equation}
F_{*}\approx\left(z_{*}\kappa_{FA}\right)^{\nu}=\frac{\kappa_{FA}}{\left[\nu^{2}\kappa_{IF}^{\nu}\kappa_{FI}\right]^{\frac{1}{\nu+1}}},\quad I_{*}\approx\frac{\kappa_{FA}}{\nu^{2}\kappa_{FI}}.\end{equation}
The ratios~(\ref{eq:combined ratios}) become\begin{equation}
p(A):p(I):p(F):p(B)\approx1:\frac{1}{\nu^{2}}:\frac{1}{\nu}:\left[\frac{\eta}{\nu^{\nu-1}}\right]^{\frac{1}{\nu+1}}.\label{eq:ratios 2}\end{equation}
This is the result of applying the RV prescription to a generic $FABI$
toy landscape in the second regime.

It remains to justify the statement of Eq.~(\ref{eq:det M sing 1}).
Below in Sec.~\ref{sub:Uniqueness-of-the} I will demonstrate that
the property~(\ref{eq:det M sing 1}) holds for a general landscape.
Here only a simple argument is presented to illustrate this property
for Eqs.~(\ref{eq:F equ 2 s})--(\ref{eq:I equ 2 s}). Using those
equations, we can express $I_{*}$ through $F_{*}$ and derive a closed
algebraic equation for $F_{*}$,\begin{equation}
F_{*}=z^{\nu}\left[\kappa_{FA}+\kappa_{FI}z^{\nu}\left(\kappa_{IB}+\kappa_{IF}F_{*}\right)^{\nu}\right]^{\nu}\equiv f(z;F_{*}).\end{equation}
The solution $F(z)$ is given by the intersection of the line $y=F$
and the curve $y=f(z;F)$ in the $y-F$ plane. The function $f(z;F)$
is convex in $F$ for $F>0,z>0$; hence, there will be a value $z=z_{*}$
for which the curve $f(z_{*};F)$ is tangent to the line $y=F$, i.e.~$f{}_{,F}(z_{*};F_{*})=1$.
This value of $z_{*}$ will then implicitly determine $F_{*}$. For
values $F\approx F_{*}$, $z\approx z_{*}$ the dependence of $F(z)$
on $z$ will exhibit the singularity behavior of the type $\sqrt{z_{*}-z}$.
To see this formally, we may expand\begin{align}
F & =f(z;F)\approx F_{*}+f_{,F}\left(F-F_{*}\right)+\frac{1}{2}f_{,FF}\left(F-F_{*}\right)^{2}\nonumber \\
 & +f_{,z}\left(z-z_{*}\right).\end{align}
Since $f_{,F}(z_{*};F_{*})=1$, we obtain \begin{equation}
F\approx F_{*}+\left.\frac{2f_{,z}}{f_{,FF}}\right|_{F_{*},z_{*}}\sqrt{z_{*}-z}.\end{equation}
This shows explicitly the singularity structure of the form~(\ref{eq:f sing}).
The determinant of the matrix $\hat{M}$ is a smooth function of $F(z)$
and $I(z)$ near $z=z_{*}$. Expressing $\det\hat{M}$ as a function
only of $F$, we obtain for $z\approx z_{*}$ the required formula~(\ref{eq:det M sing 1}),\begin{equation}
\det\hat{M}(z)\approx\left(F-F_{*}\right)\left.\frac{d}{dF}\right|_{F_{*}}\det\hat{M}\propto\sqrt{z_{*}-z}.\end{equation}

\subsection{{}``Boltzmann brains''\label{sub:The-Boltzmann-brain}}

Let us now use the same techniques to compute the relative abundance
of {}``Boltzmann brain'' observers to ordinary observers.

We need to introduce an appropriate set of generating functions. {}``Boltzmann
brains'' can be created with a fixed probability per unit 4-volume,
unlike ordinary observers who can appear only within a narrow interval
of time after creation of a given bubble. Let us therefore compare
the total number of bubbles $n_{j}$ with the total number of 4-volumes
$H_{j}^{-4}$ in bubbles of type $j$.

To be specific, let us fix $j=F$ (no Boltzmann brains can be expected
in a terminal vacuum). Consider the probability $p(n_{\text{tot}},n_{F},N_{F};k)$
of having a finite total number $n_{\text{tot}}$ of bubbles of which
$n_{F}$ are of type $F$ and $N_{F}$ is the total number of $H$-regions
of type $F$, if one starts from a single initial $H$-region of type
$k$ ($k=F,I$). The number of {}``Boltzmann brains'' in bubbles
of type $F$ is proportional to $N_{F}$ with a small proportionality
constant, $\kappa_{F}^{BB}$, which we will include at the end of
the calculation. A generating function $g$ for the probability distribution
$p$ can be defined by \begin{align}
g(z,q,r;k) & \equiv\negmedspace\sum_{n,n_{F},N_{F}\geq0}\negmedspace z^{n}q^{n_{F}}r^{N_{F}}p(n,n_{F},N_{F};k)\nonumber \\
 & \equiv\left\langle z^{n}q^{n_{F}}r^{N_{F}}\right\rangle _{n<\infty;k}.\label{eq:g FABI def BB}\end{align}
For the initial bubble types $k=F$ and $k=I$, let us denote for
brevity \begin{equation}
F(z,q,r)\equiv g(z,q,r;F),\quad I(z,q,r)\equiv g(z,q,r;I).\end{equation}
The generating functions $F$ and $I$ satisfy the following system
of equations, \begin{align}
F^{\frac{1}{\nu}} & =z\kappa_{FA}+z\kappa_{FI}I+r\kappa_{FF}F,\label{eq:F equ 1 BB}\\
I^{\frac{1}{\nu}} & =z\kappa_{IB}+zrq\kappa_{IF}F+\kappa_{II}I.\label{eq:I equ 1 BB}\end{align}
These equations differ from the analogous Eqs.~(\ref{eq:F equ 1})--(\ref{eq:I equ 1})
in that the generating parameter $q$ appears when a \emph{new} \emph{bubble}
of type $F$ is created, while the parameter $r$ appears every time
a new $H$-region of type $F$ is created, which can happen via Hubble
expansion of old $F$-bubbles as well as through nucleation of new
$F$-bubbles. 

We would like to compare the mean number of $H$-regions of type $F$
with the mean number of new bubbles of type $F$, so we compute (e.g.~starting
with $I$-bubbles)\begin{equation}
\frac{\left\langle N_{F}\right\rangle _{n;I}}{\left\langle n_{F}\right\rangle _{n;I}}=\left.\frac{\partial_{z}^{n}\partial_{r}I}{\partial_{z}^{n}\partial_{q}I}\right|_{z=0,r=1,q=1}\label{eq:BB pre ans 1}\end{equation}
and take the limit $n\rightarrow\infty$. Taking the limit as $n\rightarrow\infty$
of the ratio $\partial_{z}^{n}F_{,r}/\partial_{z}^{n}F_{,q}$ will
yield the same result, but the limit of Eq.~(\ref{eq:BB pre ans 1})
is more straightforwardly analyzed.

As before, we first obtain the equations for the derivatives $F_{,q}$,
$I_{,q}$, $F_{,r}$, $I_{,r}$ at $q=r=1$ from Eqs.~(\ref{eq:F equ 1 BB})--(\ref{eq:I equ 1 BB}).
The derivatives $F_{,q}$ and $I_{,q}$ satisfy the same equations
as before, namely Eq.~(\ref{eq:FqF equ 1 pre}), while $F_{,r}$
and $I_{,r}$ satisfy \begin{equation}
\hat{M}(z)\left[\begin{array}{c}
F_{,r}\\
I_{,r}\end{array}\right]=\left[\begin{array}{c}
\kappa_{FF}F\\
z\kappa_{IF}F\end{array}\right],\label{eq:FrF equ 1 pre}\end{equation}
whose solution is\begin{equation}
\left[\begin{array}{c}
F_{,r}\\
I_{,r}\end{array}\right]=\hat{M}^{-1}(z)\left[\begin{array}{c}
\kappa_{FF}F\\
z\kappa_{IF}F\end{array}\right].\label{eq:FrF eq 1}\end{equation}
Using the same arguments as in the previous section, we evaluate the
limit of Eq.~(\ref{eq:BB pre ans 1}),\begin{equation}
\lim_{n\rightarrow\infty}\left.\frac{\partial_{z}^{n}\partial_{r}I}{\partial_{z}^{n}\partial_{q}I}\right|_{z=0,r=1,q=1}=\nu F_{*}^{1-\frac{1}{\nu}}\kappa_{FF}+1.\label{eq:BB ans 1}\end{equation}
To analyze this simple result, we do not actually need to use the
complicated decision procedure of Sec.~\ref{sub:A-toy-landscape}.
Since $\kappa_{FF}\approx1$ and $F_{*}<1$ in any case,%
\footnote{Typically $F_{*}\ll1$ but we can do with a weaker bound $F_{*}<1$
here.%
} the value~(\ref{eq:BB ans 1}) is bounded from above by $\nu+1$,
which is not a large number. Hence, the mean total number of $H$-regions
of type $F$ is at most $\nu+1$ times larger than the mean total
number of bubbles of type $F$. (Both numbers are finite in finite
multiverses, and the relationship persists in the limit $n_{\text{tot}}\rightarrow\infty$.)

Analogous results are obtained for regions of type $I$. We simply
need to replace $F$ with $I$ and $\kappa_{FF}$ by $\kappa_{II}$
in Eq.~(\ref{eq:BB ans 1}). 

The {}``Boltzmann brains'' are created at a very small rate $\kappa_{j}^{BB}$
per $H$-region of type $j$, while ordinary observers are created
at a much larger rate per reheated 3-volume $H_{j}^{-3}$ in bubbles
of the same type. We conclude that the RV-regulated abundance of {}``Boltzmann
brains'' is always negligible compared with the abundance of ordinary
observers in bubbles of the same type.

\section{A general landscape\label{sec:A-general-landscape}}

In the previous section we performed computations in a simple toy
model of the landscape. Let us now consider a general landscape containing
$N$ vacua, labeled $j=1$, ..., $N$. The dimensionless transition
rates $\kappa_{j\rightarrow k}$ between vacua $j$ and $k$ are considered
known. The main task is to compute the RV-regulated ratio of abundances
of bubbles of kinds $j$ and $k$. We will also compare the abundances
of ordinary observers with that of Boltzmann brains.

\subsection{Bubble abundances\label{sub:Bubble-abundances-landscape}}

It is convenient to denote by $T$ the set of terminal bubble types
and to relabel the vacua such that $j=1$, ..., $N_{r}$ are the {}``recyclable''
(nonterminal) vacua. Thus, $T=\left\{ N_{r}+1,...,N\right\} $. We
start by considering the probability $p(n,\left\{ n_{j}\right\} ;k)$
of having $n_{j}$ bubbles of type $j$, with total $n=\sum_{j}n_{j}$
bubbles of all types, to the future of an initial bubble of type $k\not\in T$.
The generating function $g(z,\left\{ q_{j}\right\} ;k)$ for this
probability distribution can be defined by a straightforward generalization
of Eq.~(\ref{eq:g FABI def}),\begin{equation}
g(z,\left\{ q_{l}\right\} ;k)\equiv\sum_{n,\left\{ n_{l}\right\} \geq0}p(n,\left\{ n_{l}\right\} ;k)z^{n}\prod_{j}q_{j}^{n_{j}}.\label{eq:g general def}\end{equation}
Here $z$ is the generating parameter for $n$, and $q_{j}$ are the
generating parameters for $n_{j}$. The generating function can be
written symbolically as the average\begin{equation}
g(z,\left\{ q_{l}\right\} ;k)\equiv\left\langle z^{n}q_{1}^{n_{1}}...q_{N}^{n_{N}}\right\rangle _{n<\infty;k},\end{equation}
where the subscript $(n<\infty)$ indicates that only the events with
a finite total number of bubbles contribute to the statistical average. 

For terminal bubble types $k\in T$, the definition~(\ref{eq:g general def})
yields $g(z,\left\{ q_{j}\right\} ;k)=1$ since there are no further
bubbles to the future of terminal bubbles, thus $n=n_{j}=0$ with
probability 1.

The generating function $g(z,\left\{ q_{j}\right\} ;k)$, $k=1,...,N_{r}$
satisfies the system of $N_{r}$ nonlinear equations \begin{align}
g^{\frac{1}{\nu}}(z,\left\{ q_{j}\right\} ;k) & =\sum_{i\not\in T,i\neq k}zq_{i}\kappa_{k\rightarrow i}g(z,\left\{ q_{j}\right\} ;i)\nonumber \\
 & \quad+\sum_{i\in T}zq_{i}\kappa_{k\rightarrow i}+\kappa_{k\rightarrow k}g(z,\left\{ q_{j}\right\} ;k),\label{eq:g general equ}\end{align}
where $\nu\equiv e^{3}$ as before, and the quantities $\kappa_{k\rightarrow k}$
defined by Eq.~(\ref{eq:kappa jj def}). A derivation of Eqs.~(\ref{eq:g general equ})
will be given below in Sec.~\ref{sub:Derivation-of-Eq.}. 

The solution of Eqs.~(\ref{eq:g general equ}) can be visualized
as $N_{r}$ analytic functions $g(z,\left\{ q_{j}\right\} ;k)$, $k=1,...,N_{r}$,
of the free parameters $z$ and $q_{j}$ ($j=1,...,N$). Arguments
similar to those of Sec.~\ref{sub:A-toy-landscape} show that the
physically significant solution of Eqs.~(\ref{eq:g general equ})
is the {}``main branch'' that has the following asymptotic form
at $z\rightarrow0$,\begin{equation}
g(z;k)=z^{\nu}\left[\sum_{i\in T}q_{i}\kappa_{k\rightarrow i}\right]^{\nu}+O(z^{2\nu-1})\label{eq:main branch landscape}\end{equation}
(see also the argument at the end of Sec.~\ref{sub:Derivation-of-Eq.}).

Once the generating function $g(z,\left\{ q_{j}\right\} ;k)$ is determined,
the RV measure gives the ratio of the mean number of bubbles of type
$j$ to that of type $k$ as \begin{equation}
\frac{p(j)}{p(k)}=\lim_{n\rightarrow\infty}\left.\frac{\partial_{z}^{n}\partial_{q_{j}}g(z,\left\{ q_{i}\right\} ;i')}{\partial_{z}^{n}\partial_{q_{k}}g(z,\left\{ q_{i}\right\} ;i')}\right|_{z=0,\left\{ q_{i}\right\} =1},\label{eq:limit i j general}\end{equation}
where, for clarity, we wrote explicitly the type $i'$ of the initial
bubble. We will now use the methods developed in Sec.~\ref{sub:Bubble-abundance}
to reduce Eq.~(\ref{eq:limit i j general}) to an expression that
does not contain limits and so can be analyzed more easily. It will
then become evident that the limit~(\ref{eq:limit i j general})
is independent of $i'$.

For a fixed bubble type $j$, we first consider the derivatives $\partial_{q_{j}}g(z,\left\{ q_{i}\right\} ;k)$,
$k=1,...,N_{r}$, evaluated at $\left\{ q_{i}\right\} =1$. Let us
denote these $N_{r}$ derivatives by $h_{j}(z;k)$,\begin{equation}
h_{j}(z;k)\equiv\left.\frac{\partial}{\partial q_{j}}\right|_{q_{i}=1}g(z,\left\{ q_{i}\right\} ;k),\quad k=1,...,N_{r}.\label{eq:hjzk def}\end{equation}
These quantities are conveniently represented by an $N_{r}$-dimensional
vector, which we will denote by $\left|h_{j}(z)\right\rangle $ using
the Dirac notation (although no connection to quantum mechanics is
present here). This vector satisfies an inhomogeneous linear equation
that follows straightforwardly by taking the derivative $\partial_{q_{j}}$
at $\left\{ q_{i}\right\} =1$ of Eqs.~(\ref{eq:g general equ}).
That equation can be written in the matrix form as follows, \begin{align}
\sum_{i=1}^{N}M_{ki}(z)h_{j}(z;i) & =z\kappa_{k\rightarrow j}g(z;j),\quad k\neq j,\label{eq:linear gen 0}\\
\sum_{i=1}^{N}M_{ki}(z)g_{,q_{j}}(z;i) & =0,\quad k=j,\label{eq:linear gen 1}\end{align}
where we denoted by $g(z;j)\equiv g(z;\left\{ q_{i}\right\} =1;j)$
the solution of Eqs.~(\ref{eq:g general equ}) at $\left\{ q_{i}\right\} =1$,
written as\begin{align}
g^{\frac{1}{\nu}}(z;k) & =z\negmedspace\sum_{i\not\in T,i\neq k}\negmedspace\kappa_{k\rightarrow i}g(z;i)\nonumber \\
 & \quad+z\sum_{i\in T}\kappa_{k\rightarrow i}+\kappa_{k\rightarrow k}g(z;k),\label{eq:g equ for q1}\end{align}
while the matrix $M_{ki}(z)$ is defined by\begin{equation}
\hat{M}(z)\equiv M_{ki}(z)=\begin{cases}
\frac{1}{\nu}g^{\frac{1}{\nu}-1}(z;k)-\kappa_{k\rightarrow k}, & k=i;\\
-z\kappa_{k\rightarrow i}, & k\neq i.\end{cases}\label{eq:matrix M}\end{equation}
For convenience we rewrite Eqs.~(\ref{eq:linear gen 0})--(\ref{eq:linear gen 1})
in a more concise form,\begin{equation}
\hat{M}(z)\left|h_{j}(z)\right\rangle =\left|Q_{j}(z)\right\rangle ,\label{eq:Mh Q}\end{equation}
where $\left|Q_{j}\right\rangle $ is the vector with the components
$\left|Q_{j}\right\rangle _{i}$, $i=1,...,N_{r}$  given by the right-hand
sides of Eqs.~(\ref{eq:linear gen 0})--(\ref{eq:linear gen 1}),\begin{equation}
\left|Q_{j}(z)\right\rangle _{i}\equiv z\kappa_{i\rightarrow j}g(z;j)\left[1-\delta_{ij}\right].\label{eq:Qj def}\end{equation}
The solution of Eq.~(\ref{eq:Mh Q}) is found symbolically as\begin{equation}
\left|h_{j}(z)\right\rangle =\hat{M}^{-1}(z)\left|Q_{j}(z)\right\rangle ,\label{eq:h sol M Q}\end{equation}
provided that the inverse matrix $\hat{M}^{-1}(z)$ exists. Of course,
it is impractical to obtain the inverse matrix explicitly; but we
will never need to do that.

The next task is to determine $z$ for which the matrix $\hat{M}(z)$
remains nondegenerate, $\det\hat{M}(z)\neq0$. It will be shown in
Sec.~\ref{sub:Uniqueness-of-the} that there exists an eigenvalue
$\lambda_{0}(z)$ of $\hat{M}(z)$ such that $\lambda_{0}(z_{*})=0$
at some real $z_{*}>0$, while $\lambda_{0}(z)>0$ for $z<z_{*}$.
Moreover, the (left and right) eigenvectors corresponding to the eigenvalue
$\lambda_{0}(z)$ are always nondegenerate and can be chosen with
all positive components. At the same time, all the other eigenvalues
of $\hat{M}(z)$ remain nonzero for all $z\leq z_{*}$. It will also
be shown (see Sec.~\ref{sub:Asymptotic-behavior-of-lambda}) that
the following asymptotic expansion holds near $z=z_{*}$, \begin{equation}
\lambda_{0}(z)=c_{1}\sqrt{z_{*}-z}+O(z_{*}-z),\quad c_{1}>0.\label{eq:lambda0 asympt}\end{equation}
One can see from the small-$z$ asymptotics~(\ref{eq:main branch landscape})
and from Eq.~(\ref{eq:matrix M}) that, for small enough $z$, $\hat{M}(z)$
contains very large positive numbers on the diagonal and very small
negative numbers off the diagonal. Hence $\det\hat{M}(z)>0$ for all
sufficiently small $z$. Since the eigenvalues of $\hat{M}(z)$ remain
nonzero except for $\lambda_{0}(z)$ that first vanishes at $z=z_{*}$,
we conclude that $\det\hat{M}(z)$ remains positive for all $0<z<z_{*}$,
and that $z=z_{*}$ is the smallest positive value of $z$ for which
$\det\hat{M}(z)=0$. 

Now we restrict our attention to $0<z<z_{*}$, for which $\det\hat{M}(z)>0$
and $\hat{M}^{-1}(z)$ exists. We need to analyze the behavior of
$\hat{M}^{-1}$ near $z\approx z_{*}$. We treat $\hat{M}(z)$ as
an operator in $N_{r}$-dimensional real space $V$. The matrix $\hat{M}$
is not symmetric and may not be diagonalizable. Instead of diagonalizing
$\hat{M}(z)$, we split the space $V$ into the 1-dimensional subspace
corresponding to the smallest eigenvalue $\lambda_{0}(z)$, and into
the $\left(N_{r}-1\right)$-dimensional complement subspace $V_{1}$.
We know that the eigenvalue $\lambda_{0}(z)$ is nondegenerate. Thus,
we may symbolically write\begin{equation}
\hat{M}(z)=\lambda_{0}(z)\left|v^{0}(z)\right\rangle \left\langle u^{0}(z)\right|+\hat{M}_{1}(z),\end{equation}
where $\left|v^{0}\right\rangle $ and $\left\langle u^{0}\right|$
are the right and the left eigenvectors corresponding to $\lambda_{0}(z)$,
normalized such that\begin{equation}
\left\langle u^{0}|v^{0}\right\rangle =1,\end{equation}
and it is implied that $\hat{M}_{1}(z)$ vanishes on $\left|v^{0}\right\rangle $
and $\left\langle u^{0}\right|$ but is nonsingular on the complement
space $V_{1}$. In other words, we have\begin{align}
\hat{M}\left|v^{0}\right\rangle  & =\lambda_{0}\left|v_{0}\right\rangle ,\quad\left\langle u^{0}\right|\hat{M}=\lambda_{0}\left\langle u^{0}\right|,\\
\hat{M}_{1}\left|v^{0}\right\rangle  & =0,\quad\left\langle u^{0}\right|\hat{M}_{1}=0.\end{align}
There exists a matrix $\hat{M}_{1(V_{1})}^{-1}$ that acts as the
inverse to $\hat{M}_{1}$ when restricted to the subspace $V_{1}$
and again vanishes on $\left|v^{0}\right\rangle $ and $\left\langle u^{0}\right|$.
Using that matrix, we may write the inverse matrix $\hat{M}^{-1}$
explicitly as\begin{equation}
\hat{M}^{-1}(z)=\frac{1}{\lambda_{0}(z)}\left|v^{0}(z)\right\rangle \left\langle u^{0}(z)\right|+\hat{M}_{1(V_{1})}^{-1}(z).\label{eq:M inverse}\end{equation}
Hence, the solution~(\ref{eq:h sol M Q}) can be written as\begin{equation}
\left|h_{j}(z)\right\rangle =\frac{\left\langle u^{0}|Q_{j}\right\rangle }{\lambda_{0}(z)}\left|v^{0}\right\rangle +\hat{M}_{1(V_{1})}^{-1}\left|Q_{j}\right\rangle .\label{eq:hj lambda0 ans}\end{equation}
Now it is clear from Eq.~(\ref{eq:lambda0 asympt}) that $\left|h_{j}(z)\right\rangle $
diverges at $z=z_{*}$ as $\propto\left(z_{*}-z\right)^{-1/2}$. It
also follows that $\left|h_{j}(z)\right\rangle $ does not diverge
at any smaller real $z$. 

We can also show that $\lambda_{0}(z)$ cannot vanish at some \emph{complex}
value of $z$ that is closer to $z=0$ than $z=z_{*}$. If $\lambda_{0}(z_{*}^{\prime})=0$
with a complex-valued $z_{*}^{\prime}$, then a derivative of the
generating function, such as $\partial_{z}g(z;j)$, would diverge
at $z=z_{*}^{\prime}$. Using the definition~(\ref{eq:g general def})
of $g$ and substituting $q_{i}=1$, we find that the following sum
diverges,\begin{equation}
\left.\frac{\partial g(z;j)}{\partial z}\right|_{z=z_{*}^{\prime}}=\sum_{n\geq1}\left(z_{*}^{\prime}\right)^{n-1}np(n;j)=\infty.\label{eq:series for dgdz}\end{equation}
The sum of absolute values is not smaller than the above, and hence
also diverges:\begin{equation}
\left.\frac{\partial g(z;j)}{\partial z}\right|_{z=\left|z_{*}^{\prime}\right|}=\sum_{n\geq1}\left|z_{*}^{\prime}\right|^{n-1}np(n;j)=\infty.\end{equation}
So $\partial_{z}g$ has also a singularity at a \emph{real} value
$z=\left|z_{*}^{\prime}\right|$. As we have shown, $z=z_{*}$ is
the smallest such real-valued singularity point; hence $\left|z_{*}^{\prime}\right|\geq z_{*}$.
It follows that $z_{*}$ is equal to the radius of convergence of
the series~(\ref{eq:series for dgdz}), which is a Taylor series
for the function $\partial_{z}g$. There remains the possibility that
a singularity $z_{*}^{\prime}$ is located directly on the circle
of convergence, so that $\left|z_{*}^{\prime}\right|=z_{*}$ and $z_{*}^{\prime}=z_{*}e^{\text{i}\phi}$
with $0<\phi<2\pi$. This possibility can be excluded using the following
argument. The function $\partial_{z}g$ can have only finitely many
singularities on the circle $\left|z\right|=z_{*}$; infinitely many
singularities on the circle would indicate an accumulation point which
would be an essential singularity, i.e.~not a branch point. However,
by construction $g(z;j)$ is an algebraic function of $z$ that cannot
have singularities other than branch points. Since $g(z;j)$ is regular
at $z=z_{*}^{\prime}$, all branch points of $g$ must be of the form
$\left(z-z_{*}^{\prime}\right)^{s}$ with $s>0$. Using the explicit
formula~(\ref{eq:large-n derivative}) for the $n$-th derivative
of an analytic function with a branch cut singularity of the form
$\left(z-z_{*}^{\prime}\right)^{s}$, we find that the coefficients
$np(n;j)$ of the Taylor series~(\ref{eq:series for dgdz}) decay
at large $n$ asymptotically as \begin{align}
np(n;j) & =\frac{1}{n!}\partial_{z}^{n}\partial_{z}g\propto\frac{\Gamma(n+1-s)}{\Gamma(n+1)}\left(z_{*}^{\prime}\right)^{-n}\left(1+O(n^{-1})\right)\nonumber \\
 & \propto n^{-s}\left(z_{*}^{\prime}\right)^{-n}\left(1+O(n^{-1})\right).\label{eq:sing contrib}\end{align}
The function $g$ has a finite number of singularities $z_{*}^{\prime}=z_{*}e^{\text{i}\phi}$
on the circle $\left|z\right|=z_{*}$, and each singularity gives
a contribution of the form~(\ref{eq:sing contrib}). Hence, we may
estimate (for sufficiently large $n$, say for $n\geq n_{0}$) \begin{equation}
np(n;j)=c_{0}n^{-s}z_{*}^{-n}\left(1+O(n^{-1})\right),\quad c_{0}>0.\end{equation}
 Then the partial sum for $n\geq n_{0}$ of the series~(\ref{eq:series for dgdz})
is estimated by\begin{align}
\sum_{n\geq n_{0}}\left(z_{*}^{\prime}\right)^{n-1}np(n;j) & \approx\sum_{n\geq n_{0}}\left(z_{*}^{\prime}\right)^{n-1}c_{0}n^{-s}z_{*}^{-n}\nonumber \\
 & =\frac{c_{0}}{z_{*}}\sum_{n\geq n_{0}}n^{-s}e^{\text{i}n\phi}.\end{align}
The latter series converges for $s>0$ and $0<\phi<2\pi$. (The neglected
terms of order $n^{-1-s}$ build an absolutely convergent series and
hence introduce an arbitrarily small error into the estimate.) Therefore,
the series~(\ref{eq:series for dgdz}) also converges at $z_{*}^{\prime}\neq z_{*}$,
contradicting the assumption that other singularities exist on the
circle $\left|z\right|=z_{*}$.

We conclude from these arguments that the singularity of $g(z;j)$
nearest to $z=0$ in the complex $z$ plane is indeed at a real value
$z=z_{*}$, and all other singular points $z_{*}^{\prime}$ satisfy
the strict inequality $\left|z_{*}^{\prime}\right|>z_{*}$. The same
statement about the locations of singularities holds for the generating
functions $g(z,\left\{ q_{i}\right\} ;j)$ and hence for their derivatives
such as $\left|h_{j}(z)\right\rangle $.

To compute the final expression~(\ref{eq:limit i j general}), we
need to evaluate the $n$-th derivative $\partial_{z}^{n}$ of the
functions $h_{j}(z;i)$ at $z=0$ and for very large $n$. To this
end, we use the formula~(\ref{eq:large-n derivative}), which requires
to know the location $z=z_{*}$ of the singularity of $h_{j}(z;i)$
nearest to $z=0$. We have just found that this singularity is at
a real value $z=z_{*}$ and has the form \begin{equation}
h_{j}(z;i)\approx\frac{1}{c_{1}\sqrt{z_{*}-z}}\left\langle u^{0}|Q_{j}\right\rangle v_{i}^{0},\quad z\approx z_{*},\label{eq:hj Qj asympt ans}\end{equation}
where $v_{i}^{0}$ is the $i$-th component of the eigenvector $\left|v_{0}\right\rangle $.
The value of the proportionality constant $c_{1}$ is not required
for computing the ratios~(\ref{eq:limit i j general}). Using Eqs.~(\ref{eq:large-n derivative}),
(\ref{eq:Qj def}), and (\ref{eq:hj Qj asympt ans}), we evaluate
the limit~(\ref{eq:limit i j general}) as\begin{equation}
\frac{p(j)}{p(k)}=\negmedspace\left.\frac{\left\langle u^{0}|Q_{j}\right\rangle v_{i'}^{0}}{\left\langle u^{0}|Q_{k}\right\rangle v_{i'}^{0}}\right|_{z=z_{*}}\negmedspace=\frac{\sum_{i\neq j}u_{i}^{0}(z_{*})\kappa_{i\rightarrow j}}{\sum_{i\neq k}u_{i}^{0}(z_{*})\kappa_{i\rightarrow k}}\frac{g(z_{*};j)}{g(z_{*};k)}.\label{eq:p jk ratio ans}\end{equation}
Since the component $v_{i'}^{0}$ cancels, we find that the limit~(\ref{eq:limit i j general})
is indeed independent of the initial bubble type $i'$. We also note
that the normalization of the eigenvector $\left\langle u^{0}\right|$
is irrelevant for the ratio.

The formula~(\ref{eq:p jk ratio ans}) can be simplified further
for \emph{nonterminal} types $j,k$ if we use the relationship $\left\langle u^{0}(z_{*})\right|\hat{M}(z_{*})=0$
together with the explicit definition~(\ref{eq:matrix M}) of $\hat{M}$:\begin{align}
0 & =\sum_{i\neq j}u_{i}^{0}(z_{*})M_{ij}(z_{*})+u_{j}^{0}(z_{*})M_{jj}(z_{*})\nonumber \\
 & =-\sum_{i\neq j}u_{i}^{0}(z_{*})z_{*}\kappa_{i\rightarrow j}+u_{j}^{0}(z_{*})\left[\frac{1}{\nu}g^{\frac{1}{\nu}-1}(z_{*};j)-\kappa_{j\rightarrow j}\right].\label{eq:identity M u0}\end{align}
The last term in the square brackets in Eq.~(\ref{eq:identity M u0})
can be neglected since $\kappa_{j\rightarrow j}\approx1$ while $g(z_{*};j)\ll1$.
Hence, Eq.~(\ref{eq:p jk ratio ans}) is simplified to\begin{equation}
\frac{p(j)}{p(k)}\approx\frac{u_{j}^{0}(z_{*})}{u_{k}^{0}(z_{*})}\left[\frac{g(z_{*};j)}{g(z_{*};k)}\right]^{1/\nu}.\label{eq:p jk ratio simplified ans}\end{equation}
This is the main formula for the RV-regulated relative abundances
of bubbles of arbitrary (nonterminal) types $j$ and $k$. For a terminal
type $k\in T$, one needs to use Eq.~(\ref{eq:p jk ratio ans}) together
with $g(z;k)\equiv1$.

We note that the expression~(\ref{eq:p jk ratio simplified ans})
depends on the components $u_{i}^{0}$ of the left eigenvector of
the matrix $\hat{M}(z_{*})$ with eigenvalue $\lambda_{0}(z_{*})=0$.
Although we have been able to evaluate the limit~(\ref{eq:limit i j general})
analytically and obtained Eq.~(\ref{eq:p jk ratio simplified ans}),
the task of computing the values of $z_{*}$, $g(z_{*};j)$, and $u_{i}^{0}(z_{*})$
remains quite difficult. The outline of the required computations
is as follows: One first needs to determine $g(z;j)$ ($j=1,...,N_{r}$)
as the {}``main branch'' of the solution of Eqs.~(\ref{eq:g equ for q1})
with the small-$z$ asymptotic given by Eq.~(\ref{eq:main branch landscape}).
The functions $g(z;j)$ determine the matrix elements $M_{ij}(z)$
using Eq.~(\ref{eq:matrix M}). Then one needs to find the smallest
value $z=z_{*}>0$ such that the determinant of the matrix $\hat{M}(z)$
vanishes. Finally, one needs to compute a left eigenvector $\left\langle u^{0}(z_{*})\right|$
that corresponds to the eigenvalue $\lambda_{0}(z_{*})=0$ of the
matrix $\hat{M}(z_{*})$. The mathematical construction shown below
guarantees that $z_{*}$ exists and that $\left\langle u^{0}(z_{*})\right|$
is nondegenerate and has all positive components; the results are
independent of the normalization of $\left\langle u^{0}\right|$.
However, a brute-force numerical computation of these quantities appears
to be impossible due to the huge number $N_{r}$ of the simultaneous
equations~(\ref{eq:g equ for q1}) and to the wide range of numerical
values of the coefficients $\kappa_{i\rightarrow j}$ in a typical
landscape. Even the numerical value of $z_{*}$ is likely to be too
large to be represented efficiently in computers. In the next section
we will consider an example landscape where an analytic approximation
can be found.

\subsection{Example landscape\label{sub:Example-landscape}}

We begin with some qualitative considerations regarding the behavior
of the functions $g(z;k)$.

To determine $g(z_{*};k)$, we need to follow the main branch $g(z;k)$
as the value of $z$ is increased from $z=0$ until $\det\hat{M}(z)$
vanishes, which will determine the value $z=z_{*}$. We note that
the asymptotic form~(\ref{eq:main branch landscape}) is valid in
some range near $z=0$. Since $g(z;k)\ll1$ at those $z$, the matrix
$\hat{M}(z)$ is dominated by large positive diagonal terms $g^{\frac{1}{\nu}-1}(z;k)\delta_{kj}$.
As $z$ increases to $z_{*}$, all the functions $g(z;j)$ also increase
while $\det\hat{M}(z)$ decreases monotonically to zero (this statement
is proved rigorously in Sec.~\ref{sub:Uniqueness-of-the}). One can
visualize the changes in the matrix elements of $\hat{M}(z)$ if one
notes that the positive diagonal terms decrease with $z$ while the
negative off-diagonal terms ($M_{jk}=-z\kappa_{j\rightarrow k}$,
$j\neq k$) grow in magnitude. Eventually $\det\hat{M}(z)$ vanishes
at $z=z_{*}$ such that the off-diagonal terms become sufficiently
large at least in some rows and columns of the matrix $\hat{M}(z)$. 

Let us determine an upper bound on $z$ such that Eq.~(\ref{eq:main branch landscape})
remains a good approximation for the generating function $g(z;j)\equiv g(z,\left\{ q_{i}=1\right\} ;j)$.
This will be the case if the term $z\sum_{i\in T}\kappa_{k\rightarrow i}$
in the right hand side of Eq.~(\ref{eq:g equ for q1}) dominates
over all other terms, for every $k$. Denoting for brevity \begin{equation}
\kappa_{k\rightarrow T}\equiv\sum_{i\in T}\kappa_{k\rightarrow i}\end{equation}
the total transition probability from $k$ to all terminal vacua,
we have therefore the conditions (for every $k$)\begin{equation}
z\kappa_{k\rightarrow T}\gg g(z;k),\quad z\kappa_{k\rightarrow T}\gg z\negmedspace\sum_{i\not\in T,i\neq k}\negmedspace\kappa_{k\rightarrow i}g(z;i).\label{eq:z bound regime 1 pre}\end{equation}
These conditions are satisfied, consistently with Eq.~(\ref{eq:main branch landscape}),
if $z$ is bounded (for every $k$) simultaneously by \begin{equation}
z\ll\kappa_{k\rightarrow T}^{-1},\quad z\ll\left[\sum_{i\not\in T,i\neq k}\negmedspace\frac{\kappa_{k\rightarrow i}}{\kappa_{k\rightarrow T}}\kappa_{i\rightarrow T}^{\nu}\right]^{-\frac{1}{\nu}}.\label{eq:z bound regime 1}\end{equation}
Since typically the rate of transitions to terminal vacua is much
smaller than the rate of transitions to dS vacua, one can expect that
this regime will include $z=1$. Nevertheless, one expects that $\det\hat{M}(z)$
remains nonzero for these $z$. The detailed consideration of the
FABI model in Sec.~\ref{sub:Bubble-abundance} showed that the value
of $z_{*}$ lies outside the regime~(\ref{eq:z bound regime 1})
unless the landscape parameters are fine-tuned. In a general landscape,
one expects the analogous fine-tuning to be much stronger or even
impossible to satisfy. In other words, one expects that the functions
$g(z_{*};k)$ at least for some $k$ will violate the conditions~(\ref{eq:z bound regime 1 pre}),
although these conditions might be satisfied for other $k$.

Now we would like to estimate the value of $z$ for which the matrix
$\hat{M}(z)$ first becomes degenerate. We can use a general theorem
for estimating the eigenvalues of a matrix (Gershgorin's theorem,
see~\cite{Lancaster:1969:ToM}, chapter 7). For each of the diagonal
elements $M_{kk}$, $k=1,...,N_{r}$ one needs to draw a circle in
the complex $\lambda$ plane, centered at the diagonal element $M_{kk}$
with radius \begin{equation}
\rho_{k}\equiv\sum_{i\neq k}\left|M_{ki}\right|.\end{equation}
 Thus one obtains $N_{r}$ circles \begin{equation}
\left|\lambda-M_{kk}\right|\leq\rho_{k},\quad k=1,...,N_{r},\end{equation}
 called the Gershgorin circles. The Gershgorin theorem says that all
the eigenvalues of a matrix $M_{jk}$ are located within the union
of these circles. An elementary proof is as follows. An eigenvector
$v_{i}$ with eigenvalue $\lambda$ can be normalized such that its
component of largest absolute value is equal to 1. Let $v_{1}=1$
be this component (renumbering the components if necessary), then
it is easy to show that $\lambda$ is within a circle with center
$M_{11}$ and radius $\rho_{1}$. Namely, we start from the eigenvalue
equation, \begin{equation}
\sum_{i\neq1}M_{1i}v_{i}+M_{11}v_{1}=\lambda v_{1},\end{equation}
and use the properties $v_{1}=1$ and $\left|v_{i}\right|\leq1$ for
$i\neq1$:\begin{equation}
\left|\lambda-M_{11}\right|=\left|\sum_{i\neq1}M_{1i}v_{i}\right|\leq\sum_{i\neq1}\left|M_{1i}\right|=\rho_{1}.\end{equation}
Applying the theorem to the matrix $\hat{M}(z)$ at small $z$ such
that Eq.~(\ref{eq:main branch landscape}) is a good approximation
for the generating function $g(z;j)$, we find that the Gershgorin
circles are centered at large positive values\begin{equation}
M_{kk}=\frac{1}{\nu}g^{\frac{1}{\nu}-1}(z;k)-\kappa_{k\rightarrow k}\approx\frac{1}{\nu}\frac{1}{\left(z\kappa_{k\rightarrow T}\right)^{\nu-1}}\gg1,\end{equation}
while the radii of the circles are small, \begin{equation}
\rho_{k}=z\sum_{i\neq k,i\not\in T}\kappa_{k\rightarrow i}\ll1.\end{equation}
Hence, for small $z$ none of the circles can contain $\lambda=0$,
so the matrix $\hat{M}(z)$ is nondegenerate. As $z$ increases, the
radii $\rho_{k}$ increase while the centers $M_{kk}$ move monotonically
towards zero. The circle closest to $\lambda=0$ is the one with the
center closest to zero and the largest radius. This circle is labeled
by $k$ with largest rates $\kappa_{k\rightarrow T}$ and $\kappa_{k\rightarrow i}$.
This value of $k$ corresponds to the high-energy vacua in the landscape,
for which tunneling to any other vacua is much easier than for low-energy
vacua. Therefore the eigenvalue $\lambda(z_{*})=0$ will be located
inside the Gershgorin circle(s) centered at $M_{kk}(z_{*})$ for $k$
corresponding to high-energy vacua. The Gershgorin circle to which
the eigenvalue belongs indicates the largest component of the eigenvector.
Hence, one expects that the eigenvector $\left\langle u^{0}(z_{*})\right|$
has the largest values of its components $u_{k}^{0}$ corresponding
to the high-energy vacua $k$. The Gershgorin theorem requires that
$\lambda_{0}(z_{*})=0$ be inside the circle centered at $M_{kk}(z_{*})$
with radius $\rho_{k}(z_{*})$, i.e.\begin{equation}
M_{kk}(z_{*})=\frac{1}{\nu}g^{\frac{1}{\nu}-1}(z_{*};k)-\kappa_{k\rightarrow k}\leq z_{*}\sum_{i\neq k,i\not\in T}\kappa_{k\rightarrow i}.\end{equation}
Disregarding the terms $\kappa_{k\rightarrow k}$, which are small
in comparison with the remaining terms, and using Eq.~(\ref{eq:g equ for q1})
for $g(z_{*};k)$ we obtain the following condition,\begin{equation}
z_{*}^{\nu}\geq\frac{1}{\nu}\frac{\left[\kappa_{k\rightarrow T}+\sum_{i\neq k,i\not\in T}\kappa_{k\rightarrow i}g(z_{*};i)\right]^{1-\nu}}{\sum_{i\neq k,i\not\in T}\kappa_{k\rightarrow i}}.\end{equation}
Proceeding further requires at least an estimate of $g(z_{*};i)$
for all $i$, which is so far not available.

While we are as yet unable to obtain an explicit estimate of $z_{*}$
and $\left\langle u^{0}\right|$ for an arbitrary landscape, let us
consider a workable example that is more realistic than the FABI model.
In this example there is a single high-energy vacuum, labeled $k=1$,
and a large number of dS low-energy vacua, $k=2,...,N_{r}$, as well
as a number of terminal vacua, $k=N_{r}+1,...,N$. We assume that
the downward tunneling rates $\kappa_{1\rightarrow i}$ for $i\neq1$
are much larger than the rates $\kappa_{i\rightarrow1}$ or $\kappa_{i\rightarrow j}$
for $i,j\neq1$. Then $g(z;1)\gg g(z;i)$ for $i\neq1$, and so the
first Gershgorin circle is the one closest to $\lambda=0$. Hence
we may expect that the eigenvalue $\lambda_{0}(z)$ always belongs
to that circle. Moreover, it is likely that $z_{*}$ is within the
regime~(\ref{eq:z bound regime 1 pre}) for the low-energy vacua
($k\neq1$) but not for the high-energy vacuum $k=1$. We will now
proceed with the calculation and later check that these assumptions
are self-consistent.

If the regime~(\ref{eq:z bound regime 1 pre}) holds for every low-energy
vacuum $k\neq1$, we have \begin{equation}
g(z;k)\approx z^{\nu}\kappa_{k\rightarrow T}^{\nu},\quad k=2,...,N_{r}.\label{eq:gzk example}\end{equation}
For the high-energy vacuum $k=1$ we use Eq.~(\ref{eq:g equ for q1})
directly to find\begin{align}
g(z;1) & =\left[\kappa_{1\rightarrow T}+z\negmedspace\sum_{i\not\in T,i\neq1}\negmedspace\kappa_{1\rightarrow i}g(z;i)+\kappa_{1\rightarrow1}g(z;1)\right]^{\nu}\nonumber \\
 & \approx z^{\nu^{2}+\nu}\left[\sum_{i=2}^{N_{r}}\kappa_{1\rightarrow i}\kappa_{i\rightarrow T}^{\nu}\right]^{\nu},\label{eq:gz1 example}\end{align}
where we have neglected the terms with $\kappa_{1\rightarrow T}$
and $\kappa_{1\rightarrow1}$. The condition that the regime~(\ref{eq:z bound regime 1 pre})
holds for $k\neq1$ is\begin{equation}
\kappa_{i\rightarrow T}\gg z\negmedspace\sum_{j\neq i,j\not\in T}\negmedspace\kappa_{i\rightarrow j}g(z;j),\quad i=2,...,N_{r}.\label{eq:regime 1 example}\end{equation}

In the matrix $\hat{M}(z)$ the off-diagonal elements $M_{ij}$ for
$i\neq1$, $j\neq1$ are much smaller than all other elements. Hence,
we can approximate $\hat{M}(z)$ by a matrix that has only its first
row, first column, and the diagonal elements,\begin{align}
\hat{M}(z) & \approx\negthickspace\left(\negthickspace\begin{array}{ccccc}
M_{11} & -z\kappa_{1\rightarrow2} & -z\kappa_{1\rightarrow3} & \cdots & -z\kappa_{1\rightarrow N_{r}}\\
-z\kappa_{2\rightarrow1} & M_{22} & 0 & \cdots & 0\\
\vdots & 0 & \ddots &  & \vdots\\
-z\kappa_{N_{r}\rightarrow1} & 0 & 0 & \cdots & M_{N_{r}N_{r}}\end{array}\negthickspace\right)\negmedspace,\nonumber \\
M_{kk} & \approx\frac{1}{\nu}g^{\frac{1}{\nu}-1}(z;k).\label{eq:Mkk example}\end{align}
This approximate matrix is much easier to analyze; in particular,
its determinant and the eigenvectors can be computed in closed form.
The determinant of this matrix is \begin{equation}
\det\hat{M}(z)\approx M_{11}...M_{N_{r}N_{r}}\left[1-\sum_{i=2}^{N_{r}}\frac{z^{2}\kappa_{1\rightarrow i}\kappa_{i\rightarrow1}}{M_{11}(z)M_{ii}(z)}\right].\end{equation}
Therefore, the condition $\det\hat{M}(z_{*})=0$ can be written as\begin{equation}
M_{11}(z_{*})\approx z_{*}^{2}\negmedspace\sum_{i=2}^{N_{r}}\frac{\kappa_{1\rightarrow i}\kappa_{i\rightarrow1}}{M_{ii}(z_{*})}.\end{equation}
Using Eqs.~(\ref{eq:gzk example}), (\ref{eq:gz1 example}), and
(\ref{eq:Mkk example}), we transform this condition to\begin{equation}
z_{*}^{-\nu-\nu^{2}}\approx\nu^{2}\left[\sum_{i=2}^{N_{r}}\kappa_{1\rightarrow i}\kappa_{i\rightarrow T}^{\nu}\right]^{\nu-1}\negmedspace\sum_{i=2}^{N_{r}}\kappa_{1\rightarrow i}\kappa_{i\rightarrow1}\kappa_{i\rightarrow T}^{\nu-1}.\end{equation}
However, one does not need this value apart from checking explicitly
that the assumptions~(\ref{eq:regime 1 example}) hold.

The left eigenvector $\left\langle u^{0}(z_{*})\right|$ corresponding
to the eigenvalue 0 of the matrix $\hat{M}(z_{*})$ is found approximately
as \begin{equation}
u_{1}^{0}=1,\quad u_{k}^{0}\approx\frac{z_{*}\kappa_{1\rightarrow k}}{M_{kk}(z_{*})},\end{equation}
where the normalization $u_{1}^{0}=1$ was chosen arbitrarily for
convenience. A perturbative improvement of this approximation along
the lines of Refs.~\cite{Schwartz-Perlov:2006hi,SchwartzPerlov:2006hz,Olum:2007yk,SchwartzPerlov:2008he}
may be possible, but the precision obtained from the present approximation
is sufficient for our purposes.

The bubble abundance ratios~(\ref{eq:p jk ratio simplified ans})
for nonterminal types are then expressed as follows,\begin{align}
\frac{p(j)}{p(k)} & \approx\frac{\kappa_{1\rightarrow j}}{\kappa_{1\rightarrow k}}\frac{\kappa_{j\rightarrow T}^{\nu}}{k_{k\rightarrow T}^{\nu}},\quad j,k=2,...,N_{r},\label{eq:pjk ans example}\\
\frac{p(1)}{p(k)} & \approx\frac{1}{\nu}\frac{\sum_{i=2}^{N_{r}}\kappa_{1\rightarrow i}\kappa_{i\rightarrow T}^{\nu}}{\kappa_{1\rightarrow k}\kappa_{k\rightarrow T}^{\nu}},\quad k=2,...,N_{r}.\label{eq:p1k ans example}\end{align}
These equations are the main result of this section, yielding RV-regulated
bubble abundances for a landscape with a large number of low-energy
vacua. The formula (\ref{eq:p1k ans example}) agrees with that obtained
in Sec.~\ref{sub:Bubble-abundance} for the ratio $p(F):p(I)\approx1/\nu$
in the FABI landscape, which may be considered a special case of the
present model with $N_{r}=2$.

\subsection{{}``Boltzmann brains''\label{sub:Boltzmann-brains-landscape}}

We now investigate the abundance of {}``Boltzmann brains'' in a
general landscape, relative to the abundance of ordinary observers.

We first need to derive the equations for the suitable generating
functions, analogously to Eqs.~(\ref{eq:F equ 1 BB})--(\ref{eq:I equ 1 BB}).
Let us introduce the generating function \begin{equation}
g(z,\left\{ q_{i}\right\} ,\left\{ r_{i}\right\} ;j)\equiv\left\langle z^{n_{\text{tot}}}\prod_{i}q_{i}^{n_{i}}r_{i}^{N_{i}}\right\rangle _{n_{\text{tot}}<\infty;j},\end{equation}
where $n_{i}$ is the total number of bubbles of type $i$ and $N_{i}$
is the total number of $H$-regions of type $i$. The number of BBs
in bubbles of type $i$ is proportional to $N_{i}$ with a proportionality
constant $\kappa_{i}^{BB}$, which is the (extremely small) nucleation
rate of a BB per Hubble 4-volume $H_{i}^{-4}$.

The generating function $g(z,\left\{ q_{i}\right\} ,\left\{ r_{i}\right\} ;j)$,
which we will denote for brevity by $g(...,j)$, satisfies a system
of equations analogous to Eq.~(\ref{eq:g general equ}),\begin{equation}
g^{\frac{1}{\nu}}(...,j)=\sum_{k\neq j}\kappa_{j\rightarrow k}zq_{k}r_{k}g(...,k)+\kappa_{j\rightarrow j}r_{j}g(...,j).\label{eq:g equ with BB}\end{equation}
Let us compute the RV-regulated ratio of the number of $H$-regions
of type $j$ to the number of bubbles of the same type $j$. This
ratio is given by\begin{equation}
\lim_{n\rightarrow\infty}\frac{\left\langle N_{j}\right\rangle _{n}}{\left\langle n_{j}\right\rangle _{n}}=\lim_{n\rightarrow\infty}\left.\frac{\partial_{z}^{n}\partial_{r_{j}}g(...,i')}{\partial_{z}^{n}\partial_{q_{j}}g(...,i')}\right|_{z=0,r_{i}=1,q_{i}=1},\end{equation}
where $i'$ is the initial bubble type. To evaluate the limit, we
use the methods developed in Sec.~\ref{sub:Bubble-abundances-landscape}.
The derivatives $\partial_{q_{j}}g\equiv\left|h_{j}(z)\right\rangle $,
as defined in Eq.~(\ref{eq:hjzk def}), satisfy Eq.~(\ref{eq:Mh Q}).
The vector of derivatives\begin{equation}
\left.\frac{\partial}{\partial r_{j}}\right|_{r_{i}=1,q_{i}=1}g(...,k)\equiv\left|\rho_{j}(z)\right\rangle _{k}\end{equation}
 satisfies the linear equations that follow from Eq.~(\ref{eq:g equ with BB}),\begin{equation}
\hat{M}(z)\left|\rho_{j}(z)\right\rangle =\left|\beta_{j}(z)\right\rangle ,\label{eq:M rho beta}\end{equation}
where $\left|\beta_{j}\right\rangle $ is the vector with the components
$\left|\beta_{j}\right\rangle _{i}$, $i=1,...,N_{r}$ defined by
\begin{equation}
\left|\beta_{j}(z)\right\rangle _{i}\equiv z\kappa_{i\rightarrow j}g(z;j)\left[1-\delta_{ij}\right]+\kappa_{j\rightarrow j}g(z;j)\delta_{ij}.\label{eq:betaj def}\end{equation}
By the same considerations that lead to Eq.~(\ref{eq:p jk ratio ans}),
we now obtain\begin{equation}
\lim_{n\rightarrow\infty}\left.\frac{\partial_{z}^{n}\partial_{r_{j}}g(...,i')}{\partial_{z}^{n}\partial_{q_{j}}g(...,i')}\right|_{z=0,r_{i}=1,q_{i}=1}=\left.\frac{\left\langle u^{0}|\beta_{j}\right\rangle }{\left\langle u^{0}|Q_{j}\right\rangle }\right|_{z=z_{*}},\end{equation}
where, as before, $\left\langle u^{0}(z_{*})\right|$ is the unique
eigenvector of the matrix $\hat{M}(z_{*})$ with eigenvalue 0. (The
normalization of $\left\langle u^{0}\right|$ is again irrelevant.)
Using the definitions~(\ref{eq:Qj def}) and (\ref{eq:betaj def})
and the relationship~(\ref{eq:identity M u0}), we compute\begin{align}
\left.\frac{\left\langle u^{0}|\beta_{j}\right\rangle }{\left\langle u^{0}|Q_{j}\right\rangle }\right|_{z=z_{*}} & =\frac{u_{j}^{0}(z_{*})\kappa_{j\rightarrow j}+\sum_{i\neq j}u_{i}^{0}(z_{*})z_{*}\kappa_{i\rightarrow j}}{\sum_{i\neq j}u_{i}^{0}(z_{*})z_{*}\kappa_{i\rightarrow j}}\nonumber \\
 & =\frac{\frac{1}{\nu}g^{\frac{1}{\nu}-1}(z_{*};j)}{\frac{1}{\nu}g^{\frac{1}{\nu}-1}(z_{*};j)-\kappa_{j\rightarrow j}}.\end{align}
The last ratio is always very close to 1 since typically $g(z_{*};j)\ll1$
while $\kappa_{j\rightarrow j}\approx1$. In particular, $g(z_{*};j)\ll1$
for vacua $j$ with low-energy Hubble scale, since for those vacua
we may approximate $g(z_{*};j)$ by Eq.~(\ref{eq:main branch landscape}),
\begin{equation}
g(z_{*};j)\approx\left[z_{*}\sum_{i\in T}\kappa_{j\rightarrow i}\right]^{\nu}\ll1.\end{equation}
 Therefore, the RV-regulated ratio $N_{j}/n_{j}$ of the total number
of $H$-regions of type $j$ to the total number of bubbles of type
$j$ is never large. 

Using this result, we can now estimate the RV-regulated ratio of BBs
to ordinary observers. The number of ordinary observers per one $H$-region
of type $j$ is not precisely known but is presumably at least of
order 1 or larger, as long as bubbles of type $j$ are compatible
with life. On the other hand, the number of Boltzmann brains per $H$-region,
i.e.~within a four-volume $H_{j}^{-4}$ of spacetime, is negligibly
small. It follows that the abundance of BBs in the RV measure is always
negligible relative to the abundance of ordinary observers in the
same bubble type.

\subsection{Derivation of Eq.~(\ref{eq:g general equ})\label{sub:Derivation-of-Eq.}}

To derive Eq.~(\ref{eq:g general equ}), we need to consider the
expansion of a single initial $H$-region during one Hubble time.
Within the {}``inflation in a box'' model, an initial $H$-region
of type $j$ is split after one Hubble time $\delta t=H_{j}^{-1}$
into $\nu\equiv e^{3}$ statistically independent {}``daughter''
$H$-regions.%
\footnote{To avoid considering a non-integer number $\nu$ of daughter regions,
we may temporarily assume that $\nu$ is an integer parameter. At
the end of the derivation, we will set $\nu\equiv e^{3}\approx20.1$
in the final equations.%
} Each of the daughter $H$-regions may change its vacuum type from
$j$ to $k\neq j$ ($k=1,...,N$) with probability $\kappa_{j\rightarrow k}$.
The quantity $\kappa_{j\rightarrow j}$ was defined for convenience
by Eq.~(\ref{eq:kappa jj def}) to be the probability of \emph{not}
changing the bubble type $j$ during one Hubble time. 

The generating function $g$ is defined by Eq.~(\ref{eq:g general def}),
\begin{equation}
g(z,\left\{ q_{i}\right\} ;j)\equiv\left\langle z^{n_{\text{tot}}}q_{1}^{n_{1}}...q_{N}^{n_{N}}\right\rangle _{n_{\text{tot}}<\infty;j}\label{eq:g def average z q}\end{equation}
where $n_{i}$ is the number of bubbles of type $i$, while the notation
$\left\langle ...\right\rangle _{n_{\text{tot}}<\infty;j}$ stands
for a probabilistic average evaluated for the initial $H$-region
of type $j$ on the sub-ensemble of finite total number of bubbles
$n_{\text{tot}}$. Note that the initial bubble of type $j$ is \emph{not}
counted in $n_{i}$ or $n_{\text{tot}}$. Our goal is to obtain a
relationship between $g(z,\left\{ q_{i}\right\} ;j)$ and the generating
functions $g(z,\left\{ q_{i}\right\} ;k)$ with $k\neq j$.

To this end, we equate two expressions for the average $\left\langle z^{n_{\text{tot}}}\prod_{i}q_{i}^{n_{i}}\right\rangle _{n_{\text{tot}}<\infty;j}$.
The first expression is the left-hand side of Eq.~(\ref{eq:g def average z q}).
The second expression is found by considering the $\nu$ daughter
$H$-regions evolved out of the initial $H$-region and by using the
fact that the same average for a daughter region of type $k$ is equal
to $g(z,\left\{ q_{i}\right\} ;k)$. However, two details need to
be accounted for: First, the generating functions $g(z,\left\{ q_{i}\right\} ;k)$
evaluated for the daughter regions do not count the daughter bubbles
themselves. Second, the type $k$ of each of the daughter regions
is a random quantity. Before deriving a general relationship, let
us illustrate the procedure using an example. 

It is possible that, say, only two of the daughter regions change
their type to $k$ while all other daughter regions retain the initial
bubble type $j$. Denote temporarily by $p_{kkj...j}$ the probability
of this event; binomial combinatorics yields\begin{equation}
p_{kkj...j}=\frac{\nu!}{2!(\nu-2)!}\kappa_{j\rightarrow k}^{2}\kappa_{j\rightarrow j}^{\nu-2}.\end{equation}
 Then the average of $z^{n_{\text{tot}}}\prod_{i}q_{i}^{n_{i}}$ receives
a contribution \begin{equation}
p_{kkj...j}z^{2}q_{k}^{2}\left[g(z,\left\{ q_{i}\right\} ;k)\right]^{2}\left[g(z,\left\{ q_{i}\right\} ;j)\right]^{2-\nu}\label{eq:combinatorial example}\end{equation}
from this event. The factor $z^{2}$ describes two additional bubbles
that contribute to $n_{\text{tot}}$; the factor $q_{k}^{2}$ accounts
for two additional bubbles of type $k$; no factors of $q_{j}$ appear
since no additional bubbles of type $j$ are generated. Finally, the
powers of $g$ account for all the bubbles generated in the daughter
$H$-regions, but these generating functions do not count the daughter
$H$-regions themselves. Those daughter $H$-regions are explicitly
counted by the extra factors $z^{2}q_{k}^{2}$.

To compute the average, we need to add the contributions from all
the possible events of this kind. Since all of the daughter $H$-regions
are independent and statistically equivalent, the average $\left\langle z^{n_{\text{tot}}}\prod_{i}q_{i}^{n_{i}}\right\rangle _{n_{\text{tot}}<\infty;j}$
splits into the product of $\nu$ averages, each evaluated over a
single daughter region.

The average over a single daughter region of type $j$ has contribution
from transitions to other types $k\neq j$ and a contribution from
the event of no transition. Let us first consider a \emph{terminal}
type $k$. With probability $\kappa_{j\rightarrow k}$ a given daughter
region becomes a vacuum of type $k$. Thereafter, no more bubbles
will be nucleated inside it; the average of $\left\langle z^{n_{\text{tot}}}\prod_{i}q_{i}^{n_{i}}\right\rangle _{n_{\text{tot}}<\infty;j}$
over that daughter region is simply $zq_{k}$, meaning that there
is a total of one bubbles and only one bubble of type $k$. Hence,
the contribution of that event to the statistical average is $\kappa_{j\rightarrow k}zq_{k}$.

Now let us consider a nonterminal type $k\neq j$, $k\not\in T$.
The corresponding contribution to the average is $\kappa_{j\rightarrow k}zq_{k}g(z,\left\{ q_{i}\right\} ;k)$.
Since we have defined $g(z,\left\{ q_{i}\right\} ;k)$$\equiv1$ when
$k$ is a terminal bubble type, we may write the contribution as $\kappa_{j\rightarrow k}zq_{k}g(z,\left\{ q_{i}\right\} ;k)$
for both terminal and nonterminal types $k\neq j$.

Finally, we consider the case of $k=j$ (the daughter region retains
the original bubble type). Since no new bubbles were nucleated, the
contribution to the average is simply $\kappa_{j\rightarrow j}g(z,\left\{ q_{i}\right\} ;j)$
without any factors of $z$ or $q_{k}$. 

Putting these ingredients together, we obtain an equation for $g(z,\left\{ q_{i}\right\} ;j)$,\begin{align}
 & g(z,\left\{ q_{i}\right\} ;j)\nonumber \\
 & =\left[\sum_{k\neq j}\kappa_{j\rightarrow k}zq_{k}g(z,\left\{ q_{i}\right\} ;k)+\kappa_{j\rightarrow j}g(z,\left\{ q_{i}\right\} ;j)\right]^{\nu}.\label{eq:g equ sum power}\end{align}
This is equivalent to Eq.~(\ref{eq:g general equ}). One can also
verify that the binomial expansion of Eq.~(\ref{eq:g equ sum power})
indeed yields all the terms such as the one given in Eq.~(\ref{eq:combinatorial example}).

We can now analyze the behavior of $g(z,\left\{ q_{i}\right\} ;j)$
in the limit $z\rightarrow0$. By definition, the generating function
$g(z,\left\{ q_{i}\right\} ;j)$ is a power series in $z$ whose coefficient
at $z^{n}$ is equal to the probability of the event that a multiverse
has exactly $n$ bubbles to the future of the initial bubble $j$.
It is clear that this power series starts with the term $z^{\nu}$,
corresponding to the probability that the initial bubble $j$ expands
exactly into $\nu$ terminal bubbles, signalling the global end of
the multiverse. The probability of having fewer than $\nu$ bubbles
in the entire multiverse is equal to zero.%
\footnote{The property that there are exactly $\nu$ daughter bubbles is, of
course, an artifact of the {}``inflation in a box'' approximation.
In the actual multiverse, one has bubbles of spherical shape that
can intersect in complicated ways, so a given bubble may end in one,
two, or any other number of terminal bubbles. However, we are using
the box approximation to obtain results in the limit of very large
total number of bubbles, so we disregard the imprecision in the description
of multiverses with a very small total number of bubbles. Also, the
value of $\nu$ may be considered a variable parameter of the {}``box''
model; the final results will not be overly sensitive to the value
of $\nu$.%
} The next term of the binomial expansion is of order $z^{2\nu-1}$
since it is the product of $z^{\nu-1}$ and $g$ itself. Therefore,
the small-$z$ behavior of the generating function $g(z,\left\{ q_{i}\right\} ;j)$
must be given by Eq.~(\ref{eq:main branch landscape}). This condition,
together with analyticity in $z$, selects the unique physically relevant
solution of Eqs.~(\ref{eq:g general equ}).

\subsection{Eigenvalues of $\hat{M}(z)$\label{sub:Uniqueness-of-the}}

According to the definition~(\ref{eq:matrix M}), the matrix $\hat{M}(z)$
has positive elements on the diagonal and nonpositive elements off
the diagonal. Such a matrix can be rewritten in the form\begin{equation}
\hat{M}(z)=\mu\hat{1}-\hat{A}(z),\end{equation}
where a constant $\mu>0$ is introduced, the notation $\hat{1}$ stands
for an identity matrix, and $\hat{A}(z)$ is a suitable nonnegative
matrix, i.e.~a matrix with all nonnegative elements. For instance,
we may choose $\mu$ as the largest of the diagonal elements of $\hat{M}$.
The theory of nonnegative matrices gives powerful results for the
eigenvalues of matrices such as $\hat{M}$ and $\hat{A}$ (see e.g.~the
book~\cite{Lancaster:1969:ToM}). For the present case, the most
important are the properties of the algebraically smallest eigenvalue
of the matrix $\hat{M}$.

It will be convenient to drop temporarily the argument $z$ since
all the results of matrix theory will hold for every fixed $z$. By
the Perron-Frobenius theorem (see \cite{Lancaster:1969:ToM}, chapter
9), under the condition of irreducibility%
\footnote{The irreducibility condition means that any two recyclable vacua in
the landscape can be connected by a chain of transitions with nonzero
nucleation rates. This condition has been discussed in Refs.~\cite{Garriga:2005av,Garriga:2008ks}.
If some subset of vacua form a {}``disconnected island'' in the
landscape, such that transitions to and from the {}``island'' are
forbidden, one can regard the {}``island'' as a separate irreducible
landscape and apply the same technique to it. Hence, we consider only
irreducible landscapes in this work.%
} a nonnegative matrix $\hat{A}$ has a unique nondegenerate, real
eigenvalue $\alpha_{0}>0$ such that all the eigenvalues of $\hat{A}$
(which may be complex-valued) are located within the circle $\left|\lambda\right|\leq\alpha_{0}$
in the complex $\lambda$ plane. This {}``dominant'' eigenvalue
$\alpha_{0}$ has a corresponding (right) eigenvector $\left|v^{0}\right\rangle $
that can be chosen with all strictly positive components $v_{i}^{0}>0$.
The same property holds for the relevant left eigenvector $\left\langle u^{0}\right|$
(the matrix need not be symmetric, so the left and the right eigenvectors
do not, in general, coincide). Therefore it is possible to choose
the eigenvectors $\left\langle u^{0}\right|$ and $\left|v^{0}\right\rangle $
such that the normalization $\left\langle u^{0}|v^{0}\right\rangle =1$
holds. This normalization will be convenient for further calculations,
and so we assume that such eigenvectors have been chosen.

It follows that $\left|v^{0}\right\rangle $ and $\left\langle u^{0}\right|$
are also the right and left eigenvectors of the matrix $\hat{M}$
with the eigenvalue \begin{equation}
\lambda_{0}\equiv\mu-\alpha_{0},\end{equation}
while all the other eigenvalues of $\hat{M}$ are located within the
circle $\left|\mu-\lambda\right|\leq\alpha_{0}$ in the complex $\lambda$
plane. Since $\alpha_{0}>0$, all the other eigenvalues of $\hat{M}$
are strictly to the right (in the complex plane) of the real eigenvalue
$\lambda_{0}$. In other words, $\lambda_{0}$ is the eigenvalue of
$\hat{M}$ with the algebraically smallest real part.

Restoring now the argument $z$ of the matrix $\hat{M}$, we find
that $\hat{M}$ always has a real, nondegenerate eigenvalue $\lambda_{0}(z)$,
which is at the same time the eigenvalue with the algebraically smallest
real part among all the eigenvalues of $\hat{M}(z)$. We know that
$\det\hat{M}(z)>0$ for sufficiently small $z$; hence $\lambda_{0}(z)>0$
for those $z$. Moreover, $\det\hat{M}(z)$ will remain positive as
long as $\lambda_{0}(z)>0$, since no other eigenvalue can become
negative unless $\lambda_{0}(z)$ first becomes negative. We will
now show that $\det\hat{M}(z)$ cannot remain positive for all real
$z>0$. It will then follow by continuity of $\lambda_{0}(z)$ that
there will be a value $z_{*}$ such that $\lambda_{0}(z)>0$ for all
$0<z<z_{*}$ but $\lambda_{0}(z_{*})=0$.

We will use the property that the inverse matrix $\hat{M}^{-1}(z)$
has all positive elements as long as $\lambda_{0}(z)>0$ (equivalently
if $\alpha_{0}<\mu$). The derivation of this property is simple:
\begin{equation}
\hat{M}^{-1}={(\mu\hat{1}-\hat{A})}^{-1}=\mu^{-1}\hat{1}+\mu^{-2}\hat{A}+\mu^{-3}\hat{A}^{2}+...,\label{eq:M inv series}\end{equation}
which yields explicitly a matrix with all nonnegative elements. {[}The
matrix-valued series in Eq.~(\ref{eq:M inv series}) converges because
all the eigenvalues of $\hat{A}$ are strictly smaller than $\mu$
by absolute value.] Moreover, the irreducibility condition means that
some chain of transitions will connect every pair of recyclable vacua;
this is equivalent to saying that for any vacua $i,j$ there exists
some integer $s$ such that $\hat{A}^{s}$ has a nonzero matrix element
${(\hat{A}^{s})}_{ij}$. Hence, every matrix element of $\hat{M}^{-1}$
is strictly positive as long as $\lambda_{0}(z)>0$.

Further, we can deduce that $g(z;j)$ is a strictly increasing, real-valued
function of $z$ for those $z$ for which $\lambda_{0}(z)>0$. To
show this, we consider the vector $\left|\partial_{z}g\right\rangle $
whose components are the $N_{r}$ derivatives $\partial_{z}g(z;j)$,
$j=1,...,N_{r}$. It follows from Eq.~(\ref{eq:g general equ}) that
the vector $\left|\partial_{z}g\right\rangle $ satisfies the inhomogeneous
equation\begin{equation}
\hat{M}(z)\left|\partial_{z}g\right\rangle =\left|\zeta\right\rangle ,\label{eq:dzg equ}\end{equation}
where we denoted by $\left|\zeta\right\rangle $ the vector with the
components \begin{equation}
\zeta_{k}(z)\equiv\sum_{i\neq k}\kappa_{k\rightarrow i}g(z;i).\end{equation}
The solution of Eq.~(\ref{eq:dzg equ}) is\begin{equation}
\left|\partial_{z}g\right\rangle =\hat{M}^{-1}(z)\left|\zeta(z)\right\rangle .\label{eq:dzg sol 1}\end{equation}
Since all the matrix elements of $\hat{M}^{-1}(z)$ are positive and
all the components of $\left|\zeta\right\rangle $ are nonnegative
as long as $g(z;i)>0$, it follows that all the components of $\left|\partial_{z}g\right\rangle $
are strictly positive. Equation~(\ref{eq:main branch landscape})
shows that $g(z;i)>0$ for sufficiently small $z>0$, and it follows
that $g(z;i)$ will remain positive for all $z>0$ such that $\lambda_{0}(z)>0$.
Therefore, $g(z;i)$, $i=1$, ..., $N_{r}$ are strictly increasing
functions of $z$ for all those $z$.

Nevertheless, the functions $g(z;i)$ are bounded from above. To see
this, consider the relationship \begin{equation}
\left\langle u^{0}\right|\hat{M}=\lambda_{0}\left\langle u^{0}\right|,\end{equation}
 written in components as\begin{equation}
u_{j}^{0}(z)\left[\frac{1}{\nu}g^{\frac{1}{\nu}-1}(z;j)-\kappa_{j\rightarrow j}\right]-\sum_{i\neq j}u_{i}^{0}(z)z\kappa_{i\rightarrow j}=\lambda_{0}u_{j}^{0}.\end{equation}
 Since all the components $u_{i}^{0}$ are strictly positive (as long
as $\lambda_{0}(z)>0$), it follows that \begin{equation}
\frac{1}{\nu}g^{\frac{1}{\nu}-1}(z;j)-\kappa_{j\rightarrow j}>0\end{equation}
and hence\begin{equation}
g(z;j)<\left(\nu\kappa_{j\rightarrow j}\right)^{-\frac{\nu}{\nu-1}}<\frac{1}{\nu^{\frac{\nu}{\nu-1}}}\approx\frac{1}{\nu}.\label{eq:gz less 1 over nu}\end{equation}

Furthermore, we can show that $\lambda_{0}(z)$ monotonically decreases
as $z$ grows. This follows from the perturbation theory formula for
nondegenerate eigenvalues, which allows us to express $d\lambda_{0}/dz$
as a matrix product with normalized left and right eigenvectors, \begin{equation}
\frac{d\lambda_{0}(z)}{dz}=\left\langle u^{0}(z)\right|\frac{d\hat{M}}{dz}\left|v^{0}(z)\right\rangle .\label{eq:dlambda dz}\end{equation}
As we have just shown, $g(z;i)$ grows with growing $z$, so $d\hat{M}/dz$
is a matrix with all nonpositive elements. Since the vectors $\left\langle u^{0}\right|$
and $\left|v^{0}\right\rangle $ have strictly positive components
while at least \emph{some} matrix elements of the nonpositive matrix
$d\hat{M}/dz$ are strictly negative, we obtain the strict inequality\begin{equation}
\frac{d\lambda_{0}(z)}{dz}=\left\langle u^{0}(z)\right|\frac{d\hat{M}}{dz}\left|v^{0}(z)\right\rangle <0.\end{equation}
Similarly, we can show that $\det\hat{M}(z)$ monotonically decreases
with $z$:\begin{equation}
\frac{d}{dz}\det\hat{M}(z)=\bigl(\det\hat{M}(z)\bigr)\text{Tr}\bigl(\hat{M}^{-1}\frac{d\hat{M}}{dz}\bigr)<0\end{equation}
since it was already found that the matrix $\hat{M}^{-1}(z)$ has
all positive elements while $d\hat{M}/dz$ has all nonpositive elements.
However, the monotonic decrease alone of $\lambda_{0}(z)$ and of
$\det\hat{M}(z)$ is not yet sufficient to establish that the matrix
$\hat{M}(z)$ actually becomes singular at some finite $z$.

The results derived so far --- the monotonic behavior of $g(z;j)$
and $\lambda_{0}(z)$, the positivity of the matrix elements of $\hat{M}^{-1}$,
the bounds on $g$ --- hold for all $z$ for which $\lambda_{0}(z)>0$.
Now we will show that $\lambda_{0}(z)$ cannot remain positive for
all real $z>0$. We can rewrite Eq.~(\ref{eq:g equ for q1}) as\begin{equation}
z\sum_{i\neq k}\kappa_{k\rightarrow i}g(z;i)=g^{\frac{1}{\nu}}(z;k)-\kappa_{k\rightarrow k}g(z;k).\end{equation}
Using the property $g(z;j)>0$ and the bound~(\ref{eq:gz less 1 over nu}),
we obtain (for every $k$) an upper bound on $z$, \begin{equation}
z=\frac{g^{\frac{1}{\nu}}(z;k)-\kappa_{k\rightarrow k}g(z;k)}{\sum_{i\not\in T,i\neq k}\kappa_{k\rightarrow i}g(z;i)+\kappa_{k\rightarrow T}}<\frac{g^{\frac{1}{\nu}}(z;k)}{\kappa_{k\rightarrow T}}<\frac{\nu^{1-\nu}}{\kappa_{k\rightarrow T}}.\label{eq:z global bound}\end{equation}
In other words, no real-valued solutions of Eq.~(\ref{eq:g equ for q1})
exist for larger $z$. Let us then show that an upper bound on $z$
contradicts the assumption that $\lambda_{0}(z)>0$ for all $z$.
We know that there exists a real-valued solution branch $g(z;j)$
near $z=0$ such that $0<g(z;j)<\infty$ and $0<\partial_{z}g(z;j)<\infty$
for all those $z>0$ for which this solution branch remains real-valued.
Hence, $g(z;j)$ can be viewed as a solution of a differential equation
$\partial_{z}g(z;j)=\left|\partial_{z}g\right\rangle $ with continuous
coefficients and everywhere positive right-hand side. The solution
of such differential equations, if bounded, will exist for all $z>0$.
Indeed, if the solution $g(z;j)$ existed only up to some $z=z_{1}$,
we would have, by assumption, $\lambda_{0}(z_{1})>0$ and hence a
finite value $g(z_{1};j)>0$ and a finite derivative $\partial_{z}g(z_{1};j)>0$.
So the solution $g(z;j)$ could then be continued further to some
$z>z_{1}$. Therefore, the real-valued solution branch $g(z;j)$ must
exist for all $z>0$. This is incompatible with the bound~(\ref{eq:z global bound}).

We conclude that there exists a value $z_{*}>0$ such that $\lambda_{0}(z_{*})=0$
but $\lambda_{0}(z)>0$ for all $0<z<z_{*}$. Within the range $0<z<z_{*}$
the functions $g(z;j)$ grow monotonically but remain bounded by Eq.~(\ref{eq:gz less 1 over nu}),
while $\lambda_{0}(z)$ and $\det\hat{M}(z)$ both decrease monotonically
to zero.

\subsection{The root of $\lambda_{0}(z)$\label{sub:Asymptotic-behavior-of-lambda}}

It remains to establish that $\lambda_{0}(z)$ indeed has the form~(\ref{eq:lambda0 asympt})
near $z=z_{*}$. We again restrict our attention to the interval $0<z<z_{*}$
where $\lambda_{0}(z)>0$. For these $z$ we expand $\lambda_{0}(z)$
in Taylor series and express the value $\lambda_{0}(z_{*})\equiv0$
as\begin{equation}
0=\lambda_{0}(z_{*})=\lambda_{0}(z)+\frac{d\lambda_{0}(z)}{dz}\left(z_{*}-z\right)+O[(z_{*}-z)^{2}],\end{equation}
hence\begin{equation}
\frac{d\lambda_{0}(z)}{dz}=-\frac{\lambda_{0}(z)}{z_{*}-z}+O(z_{*}-z).\end{equation}
We then use Eq.~(\ref{eq:dlambda dz}) to express $d\lambda_{0}/dz$
in another way,\begin{align}
\frac{d\lambda_{0}(z)}{dz} & =\left\langle u^{0}(z)\right|\frac{d}{dz}\hat{M}(z)\left|v^{0}(z)\right\rangle \nonumber \\
= & \left\langle u^{0}(z)\right|\left[\sum_{i}\frac{\partial\hat{M}}{\partial g(z;i)}\partial_{z}g(z;i)+\partial_{z}\hat{M}(z)\right]\left|v^{0}(z)\right\rangle ,\end{align}
where in the second line we interpreted $\hat{M}(z)$ as a function
of $N_{r}$ variables $g(z;i)$, $i=1,...,N_{r}$, and explicitly
of $z$, in order to express $d/dz$ through $\partial/\partial g$
and $\partial/\partial z$. Using Eq.~(\ref{eq:dzg sol 1}), we then
find\begin{equation}
\frac{\lambda_{0}(z)}{z-z_{*}}\approx\left\langle u^{0}(z)\right|\negmedspace\left[\negmedspace\sum_{i,k}\frac{\partial\hat{M}}{\partial g(z;i)}\hat{M}_{ik}^{-1}\zeta_{k}+\frac{\partial\hat{M}(z)}{\partial z}\right]\negmedspace\left|v^{0}(z)\right\rangle .\label{eq:lambda rel 0}\end{equation}
Since we are only interested in the qualitative behavior of $\lambda_{0}(z)$
at $z=z_{*}$, we do not need to keep track of the complicated coefficients
in Eq.~(\ref{eq:lambda rel 0}). Near $z=z_{*}$ we have, by Eq.~(\ref{eq:M inverse}),
\begin{equation}
\hat{M}^{-1}(z)\approx\frac{1}{\lambda_{0}(z)}\left|v^{0}\right\rangle \left\langle u^{0}\right|+O(1),\end{equation}
so the dominant singular terms in Eq.~(\ref{eq:lambda rel 0}) near
$z=z_{*}$ are \begin{equation}
-\frac{\lambda_{0}(z)}{z_{*}-z}\approx\frac{C_{1}}{\lambda_{0}(z)}+O(1).\end{equation}
Therefore we obtain \begin{equation}
\lambda_{0}(z)=c_{1}\sqrt{z_{*}-z}+O(z_{*}-z).\end{equation}
The positivity of $\lambda_{0}(z)$ for $z<z_{*}$ entails $c_{1}>0$.
This concludes the derivation of Eq.~(\ref{eq:lambda0 asympt}).

\section*{Acknowledgments}

The author thanks Martin Bucher, Jaume Garriga, Andrei Linde, Misao
Sasaki, Takahiro Tanaka, Vitaly Vanchurin, and Alex Vilenkin for valuable
discussions.

\bibliographystyle{myphysrev}
\bibliography{EI2}

\end{document}